\documentclass[preprint,showpacs,amsmath,amssymb,pre,superscriptaddress,a4paper,floatfix]{revtex4}
\usepackage{dcolumn}
\usepackage{graphicx}
\usepackage{epsfig}

\begin{document}

\title{Event-by-event simulation of Einstein-Podolsky-Rosen-Bohm experiments\footnote{Found. of Phys. (in press)}}

\author{Shuang Zhao}
\affiliation{Department of Applied Physics, Zernike Institute for Advanced Materials,
University of Groningen, Nijenborgh 4, NL-9747 AG Groningen, The Netherlands}
\author{Hans De Raedt}
\email{h.a.de.raedt@rug.nl; www.compphys.net}
\affiliation{Department of Applied Physics, Zernike Institute for Advanced Materials,
University of Groningen, Nijenborgh 4, NL-9747 AG Groningen, The Netherlands}
\author{Kristel Michielsen}
\affiliation{EMBD, Vlasakker 21, B-2160 Wommelgem, Belgium}
\date{\today}

\begin{abstract}
We construct an event-based computer simulation model
of the Einstein-Podolsky-Rosen-Bohm experiments with photons.
The algorithm is a one-to-one copy of the
data gathering and analysis procedures used in real laboratory experiments.
We consider two types of experiments,
those with a source emitting photons with opposite but otherwise unpredictable
polarization and those with a source emitting photons with
fixed polarization.
In the simulation, the choice of the direction of polarization measurement
for each detection event is arbitrary.
We use three different procedures to identify pairs of photons and
compute the frequency of coincidences by analyzing experimental data and simulation data.
The model strictly satisfies Einstein's criteria of local causality, does not rely on any concept of quantum theory
and reproduces the results of quantum theory for both types of experiments.
We give a rigorous proof that the probabilistic description of the simulation model
yields the quantum theoretical expressions for the single- and two-particle expectation values.
\end{abstract}

\maketitle

\def\sumprime{\mathop{{\sum}'}}

\section{Introduction}
\label{intro}
A fundamental problem, originating from the work of Einstein, Podolsky and Rosen (EPR)~\cite{EPR35}
and reformulated by Bohm~\cite{BOHM51}
is to explain how individual detection events, registered by different detectors
in such a way that a measurement on one particle does not have a causal
effect on the result of the measurement on another particle (Einstein's
criterion of local causality), give rise to the two-particle quantum correlations that
are found in experiments~\cite{FREE72,ASPE82a,ASPE82b,TAPS94,TITT98,WEIH98,WEIH00,ROWE01,FATA04,SAKA06}.

In Einstein-Podolsky-Rosen-Bohm (EPRB) experiments, individual events are registered, correlations between
them are calculated and are found to correspond to the two-particle correlation for the singlet state.
Since, in quantum theory, the basic equation that describes individual events is not known~\cite{HOME97},
quantum theory simply cannot be used to construct a numerical algorithm to simulate the individual events. Of course, using
pseudo-random numbers we could generate events according to the probability distribution that is obtained
by solving the Schr\"odinger equation. However, the challenge is to explain how the individual events
can give rise to the two-particle correlations of the singlet state without invoking concepts of quantum theory.

The question that we address in this paper is: Given the existing experimental data
(numbers recorded during an experiment, stored on computer disks, and analyzed long after the data is taken),
that, when analyzed properly, yields expectation values which
are in good agreement with the predictions of quantum theory~\cite{FREE72,ASPE82b,TAPS94,TITT98,WEIH98,SAKA06},
is it possible to construct an event-based simulation algorithm that satisfies Einstein's criteria of local causality,
generates the same kind of data as in experiment, and is capable of reproducing {\sl exactly}
the single- and two-particle averages of quantum theory for a system of two $S=1/2$ particles?

Within the context of local realist probabilistic models,
a rigorous proof that the existence of an algorithm that describes the outcome of real EPRB experiments
cannot be excluded, has been given earlier~\cite{LARS04}.
Although such a proof is very valuable, actually finding such algorithms using local, causal processes
to generate the probability distributions of quantum theory, is another challenge.

In this paper,
we present results of a complete simulation of Aspect-type experiments using
an Einstein local, causal event-based simulation model.
An important feature in these experiments is the arbitrariness in the choice of the directions in which the polarization
will be measured, for each individual detection event~\cite{FREE72,ASPE82a,ASPE82b,TAPS94,TITT98,WEIH98}.
This feature has not been taken into account in our earlier work~\cite{RAED06c}
but is fully accounted for in the simulation procedure that we describe in this paper.

The paper is organized as follows.
In Section~\ref{sec2} we describe the experimental set-up, the data gathering method and the data
analysis procedures used in EPRB experiments with photons, closely following Ref.~\cite{WEIH98,ASPE82b}.
The  sources used in EPRB experiments emit photons with opposite but otherwise unpredictable polarization.
Each photon propagates to an observation station consisting of a polarizer and two detectors.
In accordance with quantum theory and experiments,
we expect the two-particle correlation to agree with the expression
obtained by assuming that the quantum state is a singlet.
We refer to this experimental set-up as Case I.
Inserting polarizers between the source and the
observation stations changes the pair generation procedure such that
the two photons that enter the observation stations have a fixed polarization.
In this case, the photon intensity recorded by the detectors
behind the polarizers in each observation station obeys Malus law.
We refer to this set-up as Case II.
A brief review of the quantum theoretical description of Cases I and II is given in Section~\ref{sec3}.

In real experiments, macroscopic or microscopic, we need a well-defined procedure
to decide if two detection events stem from a single system.
In real EPRB experiments with photons, the time at which the events are registered is used for this
purpose.
However, the criterion that is used to select the events
that stem from a single two-particle system is, to considerable extent, arbitrary.
In Section~\ref{sec2a}, we study this aspect by analyzing publicly available experimental data
for an EPRB experiment with photons~\cite{WEIH98}.
We present results of an analysis using three different procedures:
\begin{itemize}
\item{First, we simply divide the time interval of measurement in equally spaced bins~\cite{GILL07a}.
For each station, we determine the number of events per bin.
From this data, we compute the coincidences.
Effectively, this procedure compares the detection times at
both stations with the time of a reference clock,
using a coincidence window with a width that is equal to the bin size.}
\item{Second, we employ the criterion used in the experiment~\cite{WEIH98,WEIH00}.
We compute the coincidences of a detection event at station 1 and a detection event at station 2
by comparing the time difference of these events with a fixed time window,
that is we use relative times to determine the coincidences.}
\item{Finally, in the third procedure~\cite{WEIH00}, we first maximize the number
of coincidences by shifting by the same amount,
the detection times of station 2 relative to those of station 1,
and then use the second procedure to count the coincidences.
This two-step procedure reproduces the published results~\cite{WEIH98}.}
\end{itemize}

Our analysis shows that the first and third procedure may yield a result
that is in reasonable agreement with the prediction of quantum theory if
the bin size or coincidence window is sufficiently small.
The data obtained by the second procedure is similar except that for
a particular choice of the time window, the result is in conflict with quantum theory.
In general, these results support the idea that the idealized EPRB gedanken experiment~\cite{BELL93,SANT05,ZUKO06}
that agrees with quantum theory cannot be performed~\cite{GILL03}.

In Section~\ref{sec4} we describe an Einstein local, causal event-based computer simulation model,
based on the EPRB experiment with photons performed by Weihs {\sl et al}~\cite{WEIH98,WEIH00}.
The crucial point of the present and of our earlier work~\cite{RAED06c,RAED07a,RAED07b,RAED07c} is that
we simulate a model of the real EPRB experiments, not of the simplified, gedanken-type version
that is commonly used~\cite{BELL93,SANT05,ZUKO06}.
We give an explicit description of the algorithm to simulate the photons one by one, the observation stations containing
the polarizers and detectors, and the data analysis procedure.
The polarizers are modeled such that we reproduce the quantum theoretical results
for Case I and Case II without changing the algorithm for the polarizers, that is the
functionality of all polarizers is the same.
In contrast to the real EPRB experiments with photons~\cite{WEIH98,WEIH00}, the number of orientations per polarizer
to choose from is not limited to two.

Section~\ref{sec7} gives a rigorous analytical treatment of the
probabilistic model of our simulation algorithm and
proves that this model can reproduce the single-particle averages and the two-particle correlation
of a system of two quantum spins for Cases I and II.
This probabilistic model is identical to the one studied in Ref.~\cite{LARS04},
except for the concrete model of the time-delay mechanism.
In Section~\ref{sec5} we present our simulation data and demonstrate that there is excellent agreement with
the results obtained from quantum theory and the probabilistic model.
In Section~\ref{sec5a}, we study the effect of the time window on the frequency of coincidences and
show that the simulation model readily reproduces published experimental data, including
the statististics of the single-detection events.
A summary of our results is given in Section~\ref{sec6}.

\begin{figure}[t]
\begin{center}
\includegraphics[height=8cm]{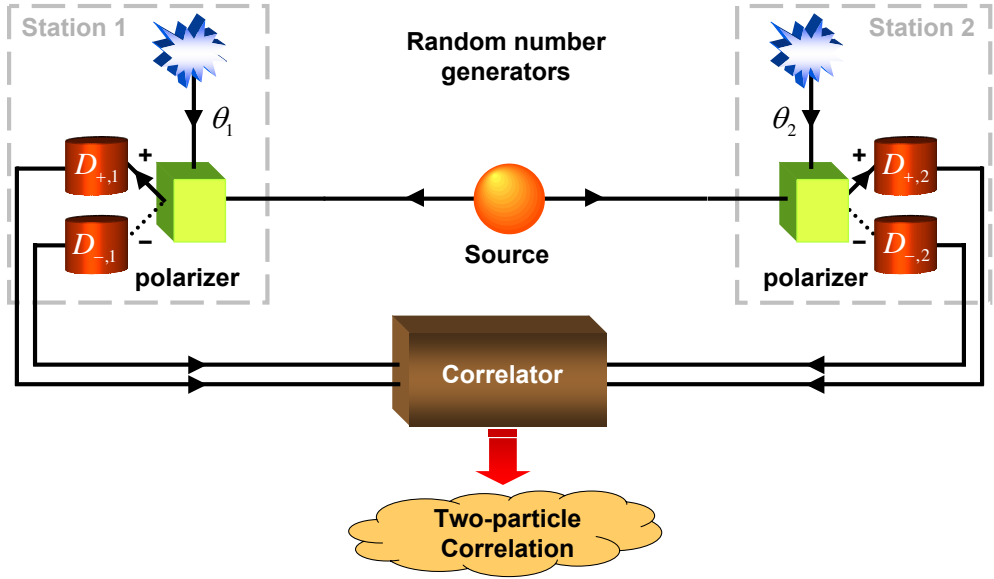}
\caption{(color online)
Case I: Schematic diagram of an EPRB experiment with randomly polarized particles.
}
\label{aspect}
\label{fig1}
\label{fig1a}
\end{center}
\end{figure}

\begin{figure}[t]
\begin{center}
\includegraphics[height=8cm]{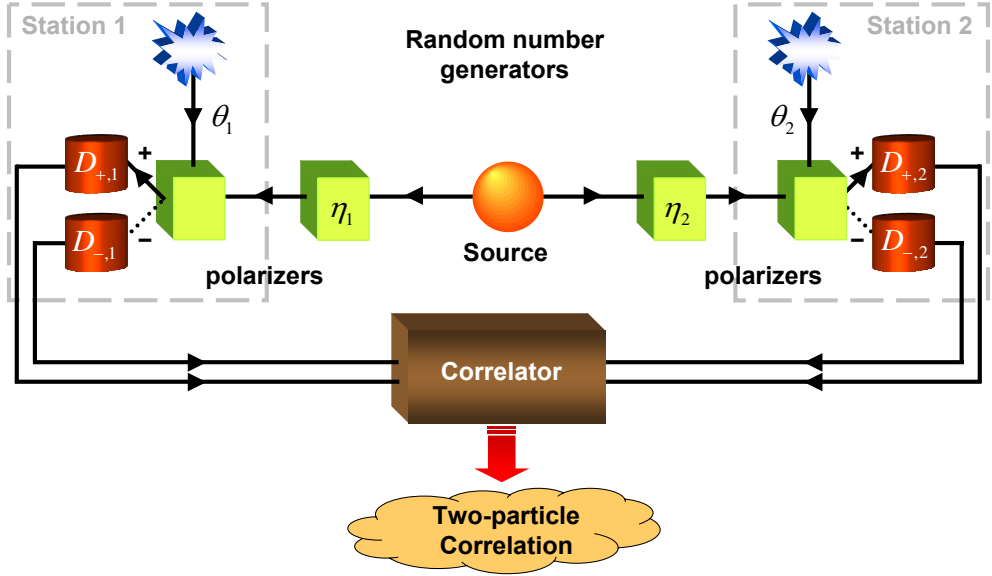}
\caption{(color online)
Case II: Schematic diagram of an EPRB experiment with particles with fixed polarization.
}
\label{fig1b}
\end{center}
\end{figure}

\section{EPRB experiment with photons}\label{sec2}

A schematic diagram of Case I is shown in Fig.~\ref{fig1a}.
A source emits pairs of photons with opposite but otherwise unpredictable polarization.
Each photon of a pair propagates to an observation station
in which it is manipulated and detected.
The two stations are separated spatially and temporally.
This arrangement prevents the observation at
station 1 (2) to have a causal effect on the
data registered at station $2$ (1).
In Case II (see Fig.~\ref{fig1b}), additional polarizers are inserted between the source
and the observation stations~\cite{ASPE82b} such that the two photons that enter the observation stations
have a fixed polarization.
We denote the orientations of these polarizers by the angles $\eta_1$ and $\eta_2$.

As the photon arrives at station $i=1,2$, it passes through a polarizer.
The orientation of the polarizer in observation station $i$ is characterized by the angle $\theta_i$,
which may be chosen at random.
As the photon leaves the polarizer, it generates a signal in one of the two detectors.
Each station has its own clock (not shown) that assigns a time-tag
to each signal generated by one of the two detectors~\cite{WEIH98,WEIH00}.
Effectively, this procedure discretizes time in intervals, the width of which is
determined by the time-tag resolution $\tau$.
In experiment, the time-tag generators are synchronized before each run~\cite{WEIH98,WEIH00}.
This procedure is necessary because in time, the clocks may become unsynchronized~\cite{WEIH98,WEIH00}.

Note that the description given earlier is only a pictorial description of real EPRB experiments with photons,
as they are carried out in a laboratory. The experimental facts are the settings of the various apparatuses
and the detection events. What happens in between activating the source and the registration of the
detection events is not, or cannot be, measured and is therefore not known. In this sense, the photon should
be regarded as an element of a model or theory for the real laboratory experiment only.

In the experiment, the firing of a detector is regarded as an event.
At the $n$th event, the data recorded on a hard disk (not shown) at station $i=1,2$
consists of $\theta_{n,i}$, $x_{n,i}=\pm 1$, specifying
which of the two detectors behind the selected polarizer fired and
the time tag $t_{n,i}$ indicating the time at which a detector fired.
Hence, the set of data collected at station $i=1,2$ during a run of $N$ events
may be written as
\begin{eqnarray}
\label{Ups}
\Upsilon_i=\left\{ {x_{n,i},t_{n,i},\theta_{n,i} \vert n =1,\ldots ,N } \right\}
.
\end{eqnarray}

Any real EPRB experiment requires some criterion to decide which detection
events are to be considered as stemming from a single two-particle system.
In EPRB-type experiments with photons, this decision is taken on the basis of coincidence in time~\cite{WEIH98,CLAU74}.
However, as discussed in Section~\ref{sec2a}, this identification procedure is not unique.
Coincidences can, for example, be identified by comparing the time differences
$\{ t_{n,1}-t_{n,2} \vert n =1,\ldots ,N \}$ with a window $W$~\cite{WEIH98}. 
In this case, for each pair of rotation angles $\alpha$ and $\beta$,
the number of coincidences between detectors $D_{x,1}$ ($x =\pm $1) at station 1 and
detectors $D_{y,2}$ ($y =\pm $1) at station 2 is given by
\begin{eqnarray}
\label{Cxy}
C_{xy}=C_{xy}(\alpha,\beta)=&
\sum_{n=1}^{N}
&\delta_{x,x_{n ,1}} \delta_{y,x_{n ,2}} \delta_{\alpha ,\theta_{n,1}}
\delta_{\beta,\theta_{n,2}}
\nonumber \\
&&\times
\Theta(W-\vert t_{n,1} -t_{n ,2}\vert)
,
\end{eqnarray}
where $\Theta (t)$ is the Heaviside step function.
The single-particle averages and correlation  between the coincidence counts
are then given by
\begin{eqnarray}
\label{Exy}
E_1(\alpha,\beta)&=&
\frac{\sum_{x,y=\pm1} x C_{xy}}{\sum_{x,y=\pm1} C_{xy}},
\nonumber \\
E_2(\alpha,\beta)&=&
\frac{\sum_{x,y=\pm1} yC_{xy}}{\sum_{x,y=\pm1} C_{xy}},
\nonumber \\
E(\alpha,\beta)&=&
\frac{\sum_{x,y=\pm1} xy C_{xy}}{\sum_{x,y=\pm1} C_{xy}}
\nonumber \\
&=&
\frac{C_{++}+C_{--}-C_{+-}-C_{-+}}{C_{++}+C_{--}+C_{+-}+C_{-+}}
,
\end{eqnarray}
where the denominators in Eq.(\ref{Exy}) are the sum of all coincidences.
In practice, the data $\{\Upsilon_1,\Upsilon_2\}$ are analyzed
long after the data have been collected~\cite{WEIH98}.
In general, the values for the coincidences
$C_{xy}(\alpha,\beta)$ depend on the time-tag resolution $\tau$
and the window $W$ used to identify the coincidences, independent of which of the three pair
identification procedures (see Section~\ref{sec2a}) is being used.

Data of EPRB experiments are often analyzed in terms of the function~\cite{WEIH98,CLAU69}
\begin{equation}
\label{eq29}
S(\alpha ,{\alpha }',\beta ,{\beta }')={E(\alpha ,\beta )-E(\alpha,{\beta }')}+{E({\alpha }',\beta )+E({\alpha }',{\beta }')},
\end{equation}
because it provides clear evidence that a quantum system is described by an entangled state.
The idea behind this is that for any product state in quantum theory, or
for the class of local realistic theories considered by Bell~\cite{BELL93}
\begin{equation}
\label{eq30}
-2\le S(\alpha, \alpha',\beta,\beta')\le 2,
\end{equation}
an inequality known as one of Bell's generalized inequalities~\cite{CLAU69}.
For later use, it is expedient to introduce the function
\begin{equation}
\label{eq31}
S(\theta )\equiv S(\alpha ,\alpha +2\theta ,\alpha +\theta ,\alpha +3\theta),
\end{equation}
where we have fixed the relation between the angles $\beta =\alpha
+\theta $, ${\alpha }'=\alpha +2\theta $, ${\beta }'=\alpha +3\theta $
through the angle $\theta $.
Assuming rotational invariance, $S(\theta )$ does not depend on $\alpha $
and we may set $\alpha=0$.

\section{Quantum Theory}\label{sec3}
\label{sec:quantum}

In this section, we give a brief account of the quantum theoretical
description of Cases I and II, strictly staying
within the axiomatic framework that quantum theory provides.

In the quantum theoretical description of Case I,
the whole system is assumed to be described by the two-particle state
\begin{eqnarray}
\label{eq7}
| \Psi \rangle &=&\frac{1}{\sqrt 2 }\left( {| H \rangle
_1 | V \rangle _2 -| V \rangle _1 | H
\rangle _2 } \right)
\nonumber \\
&=&\frac{1}{\sqrt 2 }\left( {| {HV}
\rangle -| {VH} \rangle } \right),
\end{eqnarray}
where $H$ and $V$ denote the horizontal
and vertical polarization and the subscripts refer to photon 1 and 2, respectively.
The singlet state $|\Psi\rangle$ cannot be written as a
product of single-photon states, hence it is an entangled state.

In Case II, the photons have a definite polarization when
they enter the observation station and the system is described by
the product state
\begin{equation}
\label{eq23}
|\Psi\rangle =(\cos \eta_1|H\rangle_1 +\sin \eta_1|V\rangle_1)
(\cos \eta_2 |H\rangle_2 +\sin \eta_2 |V\rangle _2).
\end{equation}

The quantum theoretical expectation $P_+(\alpha)$ ($P_-(\beta)$)
for observing a photon at the $+$ ($-$) detector behind the polarizer with orientation $\alpha$ ($\beta$)
is given in the first two rows of Table~\ref{tab1}.
The expressions for the two-particle correlation $E(\alpha,\beta)$ are given in the third row.
From Table~\ref{tab1}, it is clear that
measuring $E_1(\alpha)=P_+(\alpha)-P_-(\alpha)$, $E_2(\beta)=P_+(\beta)-P_-(\beta)$ and $E(\alpha,\beta)$
for various $\alpha$ and $\beta$
suffices to distinguish between systems in
the entangled state (Case I) or in the product state (Case II).

\begin{table}
\caption{The single- and two-particle expectation values
for the
two experiments described by the states Eqs.~(\ref{eq7}) and~(\ref{eq23}), respectively. }
\label{tab1}       
\begin{ruledtabular}
\begin{tabular}{rrr}
& Case I & Case II  \\
\noalign{\smallskip}\hline\noalign{\smallskip}
$P_+(\alpha)$ & $1/2$ & $\cos^2(\alpha-\eta_1)$\\
$P_-(\beta)$ & $1/2$ & $\sin^2(\beta-\eta_2)$\\
$E_1(\alpha)$ & $0$ & $\cos2(\alpha-\eta_1)$\\
$E_2(\beta)$ & $0$ & $\cos2(\beta-\eta_2)$\\
$E(\alpha,\beta)$ & $-\cos2(\alpha-\beta)$ & $\cos 2(\alpha-\eta_1) \cos 2(\beta-\eta_2)$\\
\end{tabular}
\end{ruledtabular}
\end{table}

In Case I, $E(\alpha ,\beta )=-\cos 2(\alpha -\beta )$
and we find
\begin{equation}
\label{Stheta}
S(\theta )=3\cos 2\theta -\cos 6\theta,
\end{equation}
which reaches its maximum value $S_{max}=\max _\theta S(\theta )=2\sqrt 2$
at $\theta =\pi /8+j\pi /2$, where $j$ is an integer number.

Analysis of the experimental data~\cite{FREE72,ASPE82b,ASPE82a,TAPS94,TITT98,WEIH98,ROWE01,FATA04,SAKA06},
yields results that are in good agreement with the expressions in Table~\ref{tab1},
leading to the conclusion that in a quantum theoretical description of Case I,
the state does not factorize, in spite of the fact that the photons are
spatially and temporally separated and do not interact.

\section{Data analysis of a real EPRB experiment with photons}\label{sec2a}

\begin{figure}[t]
\begin{center}
\mbox{
\includegraphics[width=8cm]{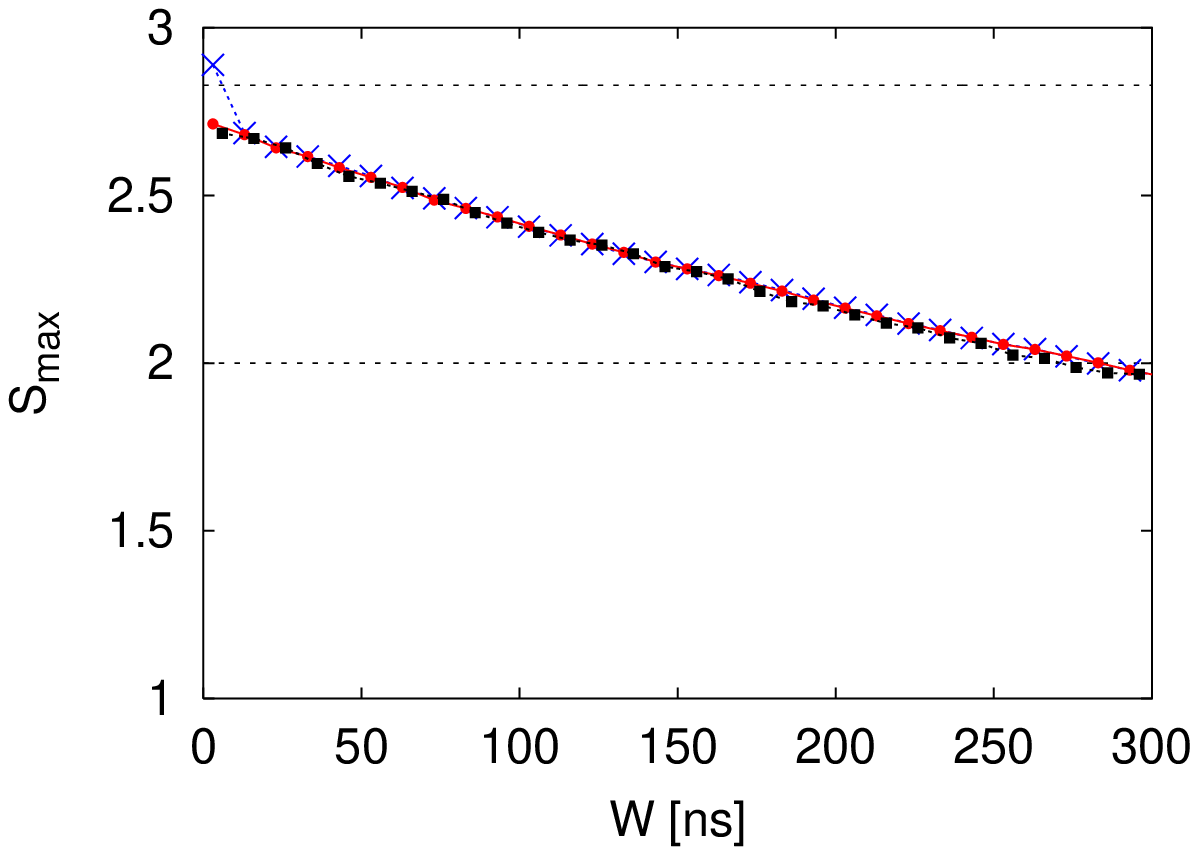}
\includegraphics[width=8cm]{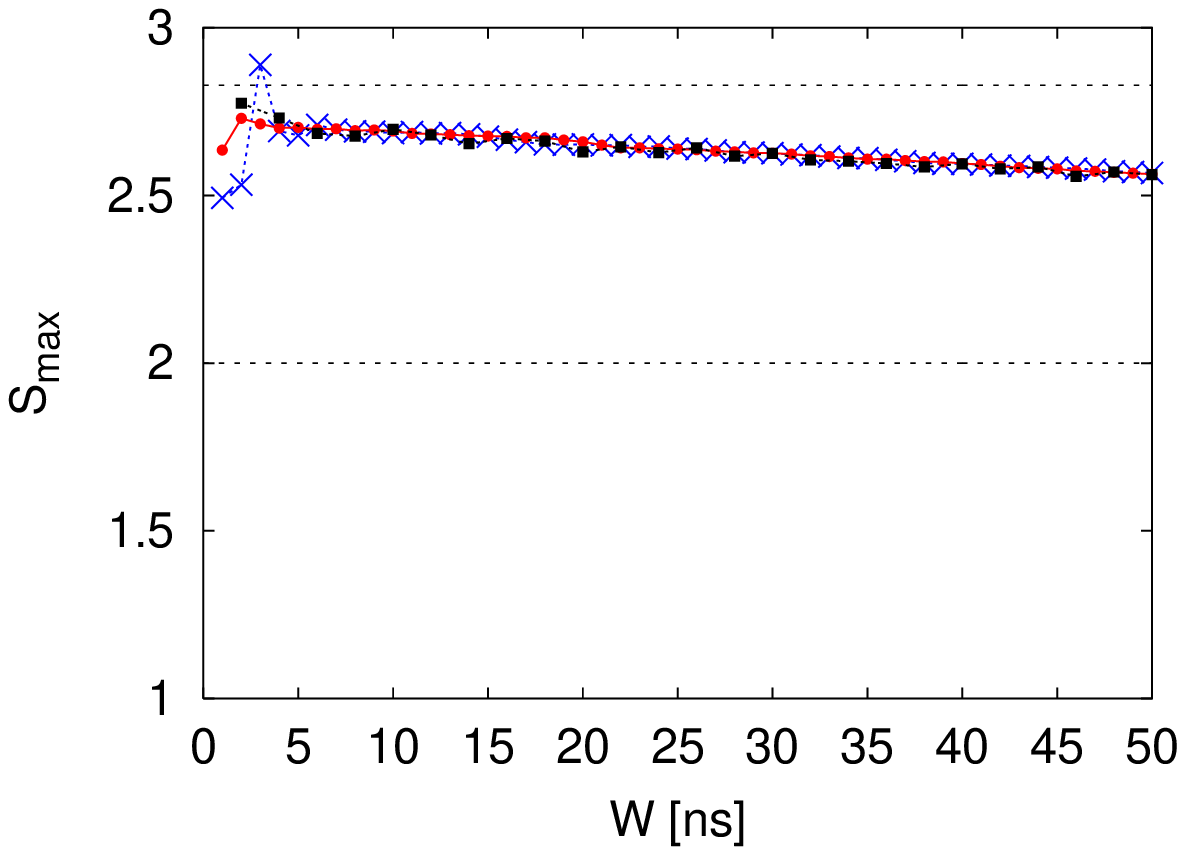}
}
\caption{(color online) Left: $S_{max}$ as a function of the time window $W$,
computed from the data sets contained in the archives
Alice.zip and Bob.zip that can be downloaded from Ref.~\cite{WEIHdownload}.
Squares (black): Data obtained by comparing the detection times
$\{t_{n,1}|n=1,\ldots,N_1=388455\}$
and
$\{t_{m,2}|m=1,\ldots,N_2=302271\}$
with a reference clock.
The maximum value of $S_{max}\approx2.78$ is found at $W=4$ ns at which
the total number of coincidences is 2010 ($\approx0.6\%$).
Crosses (blue): Results of comparing the difference of the detection times
with the time window $W$ ($\Delta=0$).
The maximum value of $S_{max}\approx2.89$ is found at $W=3$ ns
at which the total number of coincidences (with double counts removed) is 2899 ($\approx0.8\%$).
Bullets (red): Results of comparing the difference of the detection times
with the time window $W$, taking into account the time shift $\Delta=4$ ns that maximizes the
total number of coincidences, which (with double counts removed) is 13975 ($\approx4\%$) in this case.
The maximum value of $S_{max}\approx2.73$ is found at $W=2$ ns.
Dashed line at $2\sqrt 2\approx2.82 $: $S_{max}$ if the system is described by quantum theory (see Section~\ref{sec:quantum}).
Dashed line at $2$: $S_{max}$ if the system is described by the
class of models introduced by Bell~\cite{BELL93}.
Right: Same as left except for the range of $W$.
}
\label{exp2}
\end{center}
\end{figure}

\begin{figure}[t]
\begin{center}
\includegraphics[width=12cm]{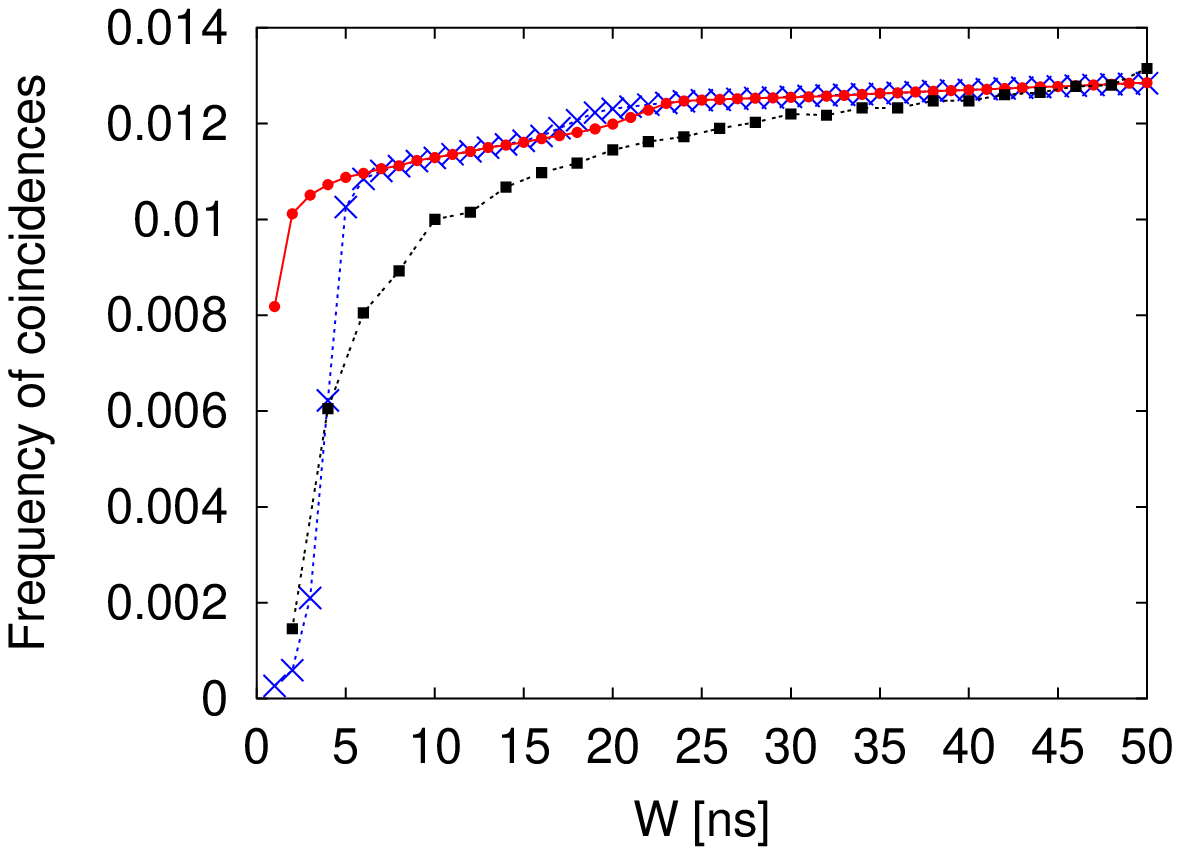}
\caption{(color online) Frequency of coincidences defined by $2(C_{++}+C_{--}+C_{+-}+C_{-+})/(N_1+N_2)$,
as a function of the time window $W$ (bin size $B=2W$).
The results were obtained by averaging the data of the four experiments
($\theta_1=0,\theta_2=\pi/8$),
($\theta_1=0,\theta_2=3\pi/8$),
($\theta_1=\pi/4,\theta_2=\pi/8$), and
($\theta_1=\pi/4,\theta_2=3\pi/8$),
contained in the archives Alice.zip and Bob.zip~\cite{WEIHdownload}.
Squares (black): Data obtained by comparing the detection times
$\{t_{n,1}|n=1,\ldots,N_1=388455\}$
and
$\{t_{m,2}|m=1,\ldots,N_2=302271\}$
with a reference clock.
Crosses (blue): Results of comparing the difference of the detection times
with the time window $W$ ($\Delta=0$).
Bullets (red): Results of comparing the difference of the detection times
with the time window $W$, taking into account the time shift $\Delta=4$ ns that maximizes the
number of coincidences.
For the values of $W$ at which the three $S_{max}$ reach their maximum
(squares: $W=4$ ns, $S_{max}\approx2.78$; crosses: $W=3$ ns, $S_{max}\approx2.89$;
bullets: $W=2$ ns, $S_{max}\approx2.73$),
the frequencies of coincidences are approximately $0.0015$,
$0.002$, and $0.01$, respectively.
}
\label{exp3}
\end{center}
\end{figure}

We analyze a data set (the archives Alice.zip and Bob.zip)
of a real EPRB experiment with photons that is publicly available~\cite{WEIHdownload}.
The archives Alice.zip and Bob.zip contain data Eq.~(\ref{Ups})
for Case I for
$\theta_1=0,\pi/4$ and $\theta_2=\pi/8,3\pi/8$.
In a real experiment, such as the one described in Ref.~\cite{WEIH98},
the number of events detected at station 1 is unlikely
to be the same as the number of events detected at station 2.
The data sets of Ref.~\cite{WEIHdownload} show that
station 1 (Alice.zip) recorded $N_1=388455$ events while
station 2 (Bob.zip) recorded  $N_2=302271$ events.
The fact that $N_1\not=N_2$ may have various reasons:
It may happen that the source emitted one instead of two photons,
one of the detectors did not respond to the arrival of a photon,
the detector fired when there was no photon, etc.
The data analysis does not account for such
possibilities: We use the data as it is, without making additional
hypotheses about unknown processes.

We need a well-defined procedure
to decide which two detection events stem from two particles that form a pair.
Here, we use three different procedures to determine these pairs.
Once the coincidences have been identified, we compute
the two-particle average and $S_{max}=\max_\theta S(\theta)$
using Eqs.(\ref{Exy}), (\ref{eq29}) and (\ref{eq31}), respectively.
In addition, we compute the frequency of coincidences, defined
by $2(C_{++}+C_{--}+C_{+-}+C_{-+})/(N_1+N_2)$.

The first procedure divides the time interval of measurement (about 10 s) in equally spaced bins
of size $B$~\cite{GILL07a}.
For each station, we use the data
$\{t_{n,1}|n=1,\ldots,N_1\}$ and
$\{t_{m,2}|m=1,\ldots,N_2\}$ to
determine the number of events per bin and compute the coincidences
by examining the content of the bins.
This procedure compares the detection times $t_{n,1}$ and $t_{m,2}$
with a reference clock, using a coincidence window $W=B/2$.
In the second procedure, we count the coincidences according to Eq.~(\ref{Cxy}).
In the third procedure, we account for the fact that in the real EPRB experiment~\cite{WEIH00},
there may be an unknown shift $\Delta$ (assumed to be constant during the experiment)
between the times $t_{n,1}$ gathered at station 1 and
the times $t_{m,2}$ recorded at station 2.
Therefore, there is some extra ambiguity in
matching the data of station 1 to the data of station 2.
A simple data processing procedure that resolves this
ambiguity consists of two steps~\cite{WEIH00}.
First, we make a histogram of the time differences
$t_{n,1}-t_{m,2}$ with a small but reasonable time resolution
(we used $0.5$ ns).
Then, we fix the value of the time-shift $\Delta$
by searching for the time difference for which
the histogram reaches its maximum.
Thus, we maximize the number of coincidences by a suitable choice of $\Delta$.
For the case at hand, we find $\Delta=4$ ns.

The results for $S_{max}$ and the frequency of coincidences,
as obtained by applying the three data analysis procedures,
are presented in Figs.~\ref{exp2} and \ref{exp3}, respectively.
Note that in general, the frequency of coincidences depends on $\alpha$ and $\beta$.
However, for the choice $\alpha=\theta_1=0,\pi/4$ and $\beta=\theta_2=\pi/8,3\pi/8$,
made in experiment~\cite{WEIH98}, by symmetry the four relevant frequencies of coincidences
are expected to be the same, hence we show their average.
As it is clear from Eq.~(\ref{Cxy}) that the width of the time window in the second and third procedure
is $2W$, we have taken $B=2W$ to facilitate the comparison.

From Fig.~\ref{exp2}, it follows that all three procedures yield a value of $S_{max}$ that significantly
exceeds the upperbound ($S_{max}=2$) of the original Bell-like models~\cite{BELL93}.
As it has been shown rigorously that the original
Bell (CHSH) inequality has to be modified if one uses Eq.~(\ref{Cxy}) to
count coincidences~\cite{LARS04}, this violation should not come as a surprise.
For $W>10$ ns and disregarding small fluctuations, the general trend is clear: $S_{max}$ decreases with $W$
and drops below the ``Bell-bound'' for $W>300$ ns.
For $W\le10$ ns, each of the three procedures yields results for $S_{max}$ that
are close to the quantum theoretical upperbound $2\sqrt{2}\approx2.83$~\cite{CIRE80}.

The procedure that maximizes the coincidence count
by varying $\Delta$ reduces the maximum value of $S_{max}$ from
a value 2.89 ($\Delta=0$) that considerably exceeds the maximum ($2\sqrt{2}$)
for the quantum system~\cite{CIRE80} to a value 2.73 (the value cited in Ref.~\cite{WEIH98})
that violates the Bell inequality~\cite{BELL93} and is less than the maximum for the quantum system.
The fact that the ``uncorrected'' data ($\Delta=0$) violate the rigorous bound for the quantum system
should not be taken as evidence that quantum theory is ``wrong'':
It merely indicates that the way in which the data of the two stations
has been grouped in two-particle events is not optimal.

Analyzing the experimental data set~\cite{WEIHdownload}
with $\Delta=4$ns and $W=2$ns (yielding the ``best'' value of $S_{max}\approx2.73$
and a total number of coincidences of 13975) gives
$E_1(0,\pi/8)=-0.073$,
$E_1(0,3\pi/8)=0.118$,
$E_1(\pi/4,\pi/8)=0.036$,
$E_1(\pi/4,3\pi/8)=-0.065$,
and
$E_2(0,\pi/8)=0.188$,
$E_2(0,3\pi/8)=0.258$,
$E_2(\pi/4,\pi/8)=0.099$,
$E_2(\pi/4,3\pi/8)=-0.147$,
significantly different from the theoretically expected value (zero, see Table~\ref{tab1}).
Disregarding the coincidence criterion,
we find $\sum_n x_{n,1}= -0.007,-0.005$
for $\alpha=0,\pi/4$
and $\sum_n x_{n,2}=-0.028,-0.024$
for $\beta=\pi/8,3\pi/4$.
Apparently, all these numbers change considerably with the station
and with the settings of the electro-optic modulators.

The results presented in Fig.~\ref{exp3},
show that for small $W$, the frequency of coincidences
depends significantly on the criterion that
is used to identify pairs of events.
The procedure that maximizes the coincidence count (procedure three)
seems to yield the most ``stable'' results.
For all three procedures, the frequency of coincidences at which $S_{max}$ reaches
its maximum is below $1\%$ (see Fig.~\ref{exp3}).
If we identify the observed frequency of coincidences with the probability
of coincidences $\gamma$ that enters the upperbound in the properly
modified Bell inequality  (see Eq.~(16) in Ref.~\cite{LARS04}), then this
theoretical upperbound is larger than 4, supporting the
idea that the experimental data~\cite{WEIH98} is not in conflict with local realism~\cite{GILL03}.

Textbook treatments of EPRB experiments assume that the correlation,
as measured in experiment, is given by~\cite{BELL93}
\begin{eqnarray}
\label{CxyBell}
C_{xy}^{(\infty)}&=&\sum_{n=1}^N\delta_{x,x_{n ,1}} \delta_{y,x_{n ,2}}
,
\end{eqnarray}
which is obtained from Eq.~(\ref{Cxy}) by taking the limit $W\rightarrow\infty$,
hence the notation $C_{xy}^{(\infty)}$.
Although the limit $W\rightarrow\infty$ defines a valid theoretical model, there is no reason
why this model should have any bearing on the real EPRB experiments with photons as they
have been performed so far.
An argument that might justify taking the limit $W\rightarrow\infty$ is
the hypothesis that for ideal experiments, the value of $W$ should
not matter. However, as our analysis of the experimental data shows,
to make contact to quantum theory, one has to reduce (not increase) $W$~\cite{WEIH98,WEIH00}.
Thus, in real EPRB experiments with photons, the window $W$ matters~\cite{WEIH98,WEIH00}.
The details of the criterion that is used to decide which two events
correspond to the observation of a single two-particle system seem to be
of secondary importance.

As it is relatively easy to reproduce the results of
quantum theory in the regime of small $W$~\cite{RAED06c},
and as keeping $W$ arbitrary does not render the mathematics more complicated,
there really is no point of studying the simplified model defined by Eq.~(\ref{CxyBell}):
We may always consider the limiting case $W\rightarrow\infty$ afterwards.

\section{Simulation model}\label{sec4}
\label{sec:computer}

We now take up the main challenge, the construction
of Einstein-local, causal processes that generate the data sets Eq.~(\ref{Ups}) such that
they reproduce the results of quantum theory, summarized in Table~\ref{tab1}.
A concrete simulation model of the EPRB experiments sketched in Figs.~\ref{fig1} and \ref{fig1b} requires
a specification of the information carried by the particles,
of the algorithm that simulates the source and
the observation stations, and of the procedure to analyze the data.
From the specification of the algorithm, it will be clear that it complies with
Einstein's criterion of local causality on the ontological level: Once the particles
leave the source, an action at observation station 1 (2) can, in no way,
have a causal effect on the outcome of the measurement at observation
station 2 (1).

\subsection{Source and particles}
The source emits particles that carry a vector
${\bf S}_{n,i}=(\cos(\xi_{n}+(i-1)\pi/2) ,\sin(\xi_{n}+(i-1)\pi/2))$,
representing the polarization of the photons.
The ``polarization state'' of a particle is completely characterized by $\xi _{n}$,
which is distributed uniformly over the whole interval $[0,2\pi[$.
We use uniform pseudo-random numbers
to mimic the apparent unpredictability of the experimental data.
However, from the description of the algorithm, it trivially follows that
instead of uniform pseudo-random number generators, simple counters that sample
the intervals $[0,2\pi[$ in a systematic, but uniform, manner might be employed
as well. This is akin to performing integrals by the trapezium rule instead
of by Monte Carlo sampling.
The source thus emits two particles with mutually orthogonal, random polarization.

In Case II we change the unpredictable polarization state of the particles
to a fixed polarization state by placing polarizers in between the source and each observation station.
These polarizers have one input and one output channel and their orientations are characterized by
the angles $\eta_1$ and $\eta_2$.

\subsection{Observation stations}

Prior to collecting data,
we fix the number $M$ of different polarization directions ($M=2$ in the experiment of Ref.~\onlinecite{WEIH98}).
We use $2M$ random numbers to fill the arrays
$(\alpha_{1} ,...,\alpha_{M})$ and $(\beta_{1} ,...,\beta_{M})$.
Before (or after) the $n$th pair leaves the source,
we use two uniform random numbers $1\le m,m'\le M$
to select the angles $\theta_{n,1} =\alpha_m$ and $\theta_{n,2} =\beta _{{m}'}$.
In practice, we use two different pseudo-random number generators for observation stations 1 and 2,
but we have never seen any statistically significant effect of using the same one for both observation stations.

\subsection{Polarizer}

We make the hypothesis that in laboratory EPRB experiments with photons the various polarizers
are interchangeable. Therefore, the algorithm to simulate the two polarizers
in the observation stations should be identical.
Evidently, for the present purpose,
if we switch from Case I to Case II, it is not
permitted to change the algorithm for the polarizer.
This also holds for the polarizers placed in between the source and the observation stations.

The input-output relation of a polarizer is rather simple: For each input
event, the algorithm maps the input vector \textbf{S} onto a single output
bit $x$. The value of the output bit depends on the orientation of the
polarizer ${\bf a}=(\cos\alpha,\sin\alpha)$. According to Malus law, for fixed
${\bf S}=(\cos\xi,\sin\xi)$ and fixed ${\bf a}$, the bits $x_{n}$ are to be generated such
that
\begin{equation}
\label{eq25}
\mathop {\lim }\limits_{N\to \infty }
\frac{1}{N}\sum\limits_{n=1}^N {x_n }
=\cos2(\xi-\alpha),
\end{equation}
with probability one.
If the input vectors \textbf{S} are distributed
uniformly over the unit circle, the sequence of output bits should satisfy
\begin{equation}
\label{eq26}
\mathop {\lim }\limits_{N\to \infty } \frac{1}{N}\sum\limits_{n=1}^N {x_n }
=0,
\end{equation}
with probability one,
independent of the orientation \textbf{a} of the polarizer.

The model for a polarizer is defined by the rule
\begin{eqnarray}
\label{sg1}
x_{n,i}=\left\{
\begin{array}{lll}
+1 & \mbox{if} & r_n\le \cos^2(\xi_n-\alpha)\\
-1 & \mbox{if} & r_n > \cos^2(\xi_n-\alpha)
\end{array}
\right.
,
\end{eqnarray}
where $0< r_n<1$ are uniform pseudo-random numbers.
The polarizer sends a particle with polarization ${\bf S}_{n,i}=(\cos\alpha ,\sin\alpha )$ or
${\bf S}_{n,i}=(-\sin\alpha ,\cos\alpha )$ through its output channel $+1$ or $-1$, respectively.
It is easy to see that for fixed $\xi_n=\xi$ and $\alpha$, this algorithm generates
events such that $2\langle x_{n,i}\rangle-1=\cos^2(\xi-\alpha)$,
where $\langle X\rangle$ denotes the average over many realizations of the
variables $r_n$ and $\xi_n$.
In this case, the input-output relation of the simulation model
agrees with Malus law Eq.~(\ref{eq25}).
On the other hand, if $\xi_n$ is distributed uniformly over the interval $[0,2\pi[$,
we have $\langle x_{n,i}\rangle=0$, in agreement with Eq.~(\ref{eq26}).
It is at this point, the model for the polarizer, that the simulation model
differs from the one used in Ref.~\onlinecite{RAED06c}: The model
of the polarizer used in Ref.~\onlinecite{RAED06c} can reproduce
the correlation of the singlet state but cannot reproduce Malus law.

In Case II we discard particles with polarization $\eta_1 +\pi/2$ ($\eta_2 +\pi/2$)
that leave the polarizers placed in between the source and observation station 1 (2).

\subsection{Time delay}
In our model, the time delay ${t}_{n ,i} $ for a particle
is assumed to be distributed uniformly over the interval $[t_{0}, t_{0}+T]$.
In practice, we use uniform pseudo-random numbers
to generate ${t}_{n ,i}$.
As in the case of the angles $\xi_{n}$, the random choice of ${t}_{n ,i}$
is merely convenient, not essential.
From Eq.(\ref{Cxy}), it follows that only differences of time delays matter.
Hence, we may put $t_0=0$.
The time-tag for the event $n$ is then $t_{n,i}\in[0,T]$.

There are not many reasonable options to choose the functional dependence of $T$.
Assuming that the particle ``knows'' its own direction
and that of the polarizer only, $T$ should be
a function of the relative angle only.
Furthermore, consistency with classical electrodynamics requires that
functions that depend on the polarization have period $\pi$~\cite{BORN64}.
Thus, we must have
$T(\xi _{n } -\theta_1)=F( ({\bf S}_{n ,1} \cdot {\bf a})^2)$
and, similarly,
$T(\xi _{n } -\theta_2)=F( ({\bf S}_{n ,2} \cdot {\bf b})^2)$,
where ${\bf b}=(\cos\beta,\sin\beta)$.
We found that $T(x)=T_0 |\sin 2x|^d$ yields the desired results~\cite{RAED06c}.
Here, $T_0 =\max_\theta T(\theta)$ is the maximum time delay
and defines the unit of time, used in the simulation.
In our numerical work, we set $T_0=1$.

\bigskip
\subsection{Data analysis}

For fixed $N$, the algorithm described earlier generates the data sets $\Upsilon_i$,
just as experiment does. In order to count the coincidences, we choose a time-tag resolution
$0<\tau<T_0$ and a coincidence window $W\ge \tau$.
We clear all the coincidence counts
$C_{xy}(\alpha_m,\beta_{m^\prime})$ for all $x,y=\pm 1$ and $m,m^\prime=1,\dots,M$.
Then, we make a loop over all events.
To count the coincidences, we first compute the discretized time tags $k_{n ,i} =\lceil t_{n ,i}/ \tau\rceil $
for all events in both data sets.
Here $\lceil{x}\rceil$ denotes the smallest integer that is larger or equal to $x$, that is
$\lceil{x}\rceil-1<x\le\lceil{x}\rceil$.
According to the procedure adopted in the experiment~\cite{WEIH98},
an entangled photon pair is observed if and only if
$\left| {k_{n,1} -k_{n,2} } \right|<k=\lceil{W/\tau}\rceil$.
Thus, if $\left| {k_{n,1} -k_{n,2} } \right|<k$,
we increment the count $C_{x_{n,1},x_{n,2}} (\alpha_m,\beta_{m^\prime} )$.

We emphasize that the simulation procedure counts
all events that, according to the same criterion as the one employed in experiment,
correspond to the detection of two-particle systems.
Note that in our simulation model, the three different methods that we used to analyze
the experimental data (see Section~\ref{sec2a}) give identical results.

\section{Probabilistic treatment}\label{sec7}

Let us assume that we can analyze our simulation model, described in Section~\ref{sec4},
by replacing the deterministic sequence of pseudo-random numbers by the mathematical
concept of independent random variables, as defined in the (Kolmogorov) theory of probability~\cite{GRIM95,JAYN03}.
Under this assumption, each event constitutes a Bernouilli trial~\cite{GRIM95,JAYN03}
and we can readily obtain analytical expressions for
the expectation values that we compute with the simulation model.

This section serves three purposes.
First, it provides a rigorous proof that for up to first order in $W$ and for $d=4$, the probabilistic description
of the simulation model {\sl exactly} reproduces the single particle averages and the two-particle correlations
of quantum theory for the system under consideration.
Second, it illustrates how the presence of the time-window introduces correlations
that cannot be described by the original Bell-like ``hidden-variable'' models~\cite{LARS04}.
Third, it reveals a few hidden assumptions that are implicit in the
derivation of the specific, factorized form of the two-particle correlation
that is essential to Bell's work.

As explained in Section~\ref{sec2}, real EPRB experiments with photons produce the data sets
\begin{eqnarray}
\label{Ups2}
\Upsilon_i=\left\{ {x_{n,i} =\pm 1,t_{n,i},\theta_{n,i} \vert n =1,\ldots ,N } \right\}
.
\end{eqnarray}

Let us {\sl assume} that there exists a probability, denoted by $P(x_1,x_2,t_1,t_2|\alpha,\beta)$, to
observe the data $\{x_1,t_1,\theta_{1}=\alpha\}$ and $\{x_2,t_2,\theta_{2}=\beta\}$ at station 1 and 2, respectively.
Notice that we assume, unlike in the computer simulation model where $\theta_{n,i}$ may change
with each event $n$ but as in the case of quantum theory, that $\alpha$ and $\beta$ are fixed.
The mathematical expectation of the coincidences $C_{xy}$ (see Eq.~(\ref{Cxy})), that is the average computed with
$P(x_1,x_2,t_1,t_2|\alpha,\beta)$, is given by
\begin{widetext}
\begin{equation}
\label{pxx1}
\langle C_{xy}\rangle \equiv N
\int_{-\infty}^{+\infty}dt_1 \int_{-\infty}^{+\infty} dt_2\,P(x,y,t_1,t_2|\alpha,\beta)\Theta(W-|t_1 -t_2|)
.
\end{equation}
Once we know $\langle C_{xy}\rangle$, the mathematical expectation
of the single-particle counts and two-particle coincidences follow from
\begin{eqnarray}
\label{pxx2}
E_1(\alpha,\beta,W)=
\frac{\sum_{x,y=\pm1} x\langle C_{xy}\rangle}{\sum_{x,y=\pm1} \langle C_{xy}\rangle},
\nonumber \\
E_2(\alpha,\beta,W)=
\frac{\sum_{x,y=\pm1} y\langle C_{xy}\rangle}{\sum_{x,y=\pm1} \langle C_{xy}\rangle},
\nonumber \\
E(\alpha,\beta,W)=
\frac{\sum_{x,y=\pm1} xy\langle C_{xy}\rangle}{\sum_{x,y=\pm1} \langle C_{xy}\rangle}
.
\end{eqnarray}

As a first step, let us express the probability for observing the data $\{x_1,x_2,t_1,t_2\}$
as an integral over the mutually exclusive events $\xi_1,\xi_2$.
According to the rules of probability theory~\cite{GRIM95,JAYN03}, we have
\begin{equation}
\label{pepr0}
P(x_1,x_2,t_1,t_2|\alpha,\beta)=
\frac{1}{4\pi^2}\int_0^{2\pi} \int_0^{2\pi}
P(x_1,x_2,t_1,t_2|\alpha,\beta,\xi_1,\xi_2) P(\xi_1,\xi_2|\alpha,\beta)
d\xi_1 d\xi_2
,
\end{equation}
where $\xi_1$ and $\xi_2$ denote the two-dimensional unit vectors, representing the polarization.
Starting from the exact representation Eq.~(\ref{pepr0}), we now {\sl assume}
that in the probabilistic version of our simulation model, for each event, the
values of $\{x_1,x_2,t_1,t_2\}$ are independent of each other and
that the values of $\{x_1,t_1\}$ ($\{x_2,t_2\}$) are also independent of
$\beta$ and $\xi_2$ ($\alpha$ and $\xi_1$)). Thus, we may write
\begin{eqnarray}
\label{pepr1}
P(x_1,x_2,t_1,t_2|\alpha,\beta)
&=&
\frac{1}{4\pi^2}\int_0^{2\pi} \int_0^{2\pi}
P(x_1,t_1|x_2,t_2,\alpha,\beta,\xi_1,\xi_2)
P(x_2,t_2|\alpha,\beta,\xi_1,\xi_2)
\nonumber \\&&\hskip 8cm\times
P(\xi_1,\xi_2|\alpha,\beta)
d\xi_1 d\xi_2
\nonumber \\
&=&
\frac{1}{4\pi^2}\int_0^{2\pi} \int_0^{2\pi}
P(x_1,t_1|\alpha,\xi_1)
P(x_2,t_2|\beta,\xi_2)
P(\xi_1,\xi_2|\alpha,\beta)
d\xi_1 d\xi_2
\nonumber \\
&=&
\frac{1}{4\pi^2}\int_0^{2\pi} \int_0^{2\pi}
P(x_1|\alpha,\xi_1)
P(t_1|\alpha,\xi_1)
P(x_2|\beta,\xi_2)
P(t_2|\beta,\xi_2)
\nonumber \\&&\hskip 8cm\times
P(\xi_1,\xi_2|\alpha,\beta)
d\xi_1 d\xi_2
\nonumber \\
&=&
\frac{1}{4\pi^2}\int_0^{2\pi} \int_0^{2\pi}
P(x_1|\alpha,\xi_1)
P(t_1|\alpha,\xi_1)
P(x_2|\beta,\xi_2)
P(t_2|\beta,\xi_2)
\nonumber \\&&\hskip 8cm\times
P(\xi_1,\xi_2)
d\xi_1 d\xi_2
,
\end{eqnarray}
where, in the last step, we {\sl assumed} that the values of
$\xi_1$ and $\xi_2$ are independent of $\alpha$ or $\beta$.
With the three assumptions made so far,
Eq.~(\ref{pepr1}) gives the exact probabilistic description of our simulation model.
It is of interest to note that Eq.~(\ref{pepr1}) can be derived directly from
the description of the algorithm, without recourse to probability theory,
by letting the number of events in the discrete sums approach infinity.

The mathematical structure of Eq.~(\ref{pepr1}) is the same as the one
that is used in the derivation of Bell's results and if we would go ahead in the same way,
our model also cannot produce the correlation of the singlet state.
However, the real factual situation in the experiment~\cite{WEIH98} is different:
The events are selected using a time window $W$ that the experimenters try to make as small as possible~\cite{WEIH00}.
Accounting for the time window, that is multiplying Eq.~(\ref{pepr1}) by the step function
and integrating over all $t_1$ and $t_2$, the expression for the probability for observing
the event $(x_1,x_2)$ reads
\begin{eqnarray}
\label{pepr2}
P(x_1,x_2|\alpha,\beta)
&=&
\frac{
\int_0^{2\pi} \int_0^{2\pi}
P(x_1|\alpha,\xi_1)P(x_2|\beta,\xi_2)
w(\alpha,\beta,\xi_1,\xi_2,W)
P(\xi_1,\xi_2)
d\xi_1 d\xi_2
}{
\sum_{x_1,x_2=\pm1}
\int_0^{2\pi} \int_0^{2\pi}
P(x_1|\alpha,\xi_1)
P(x_2|\beta,\xi_2)
w(\alpha,\beta,\xi_1,\xi_2,W)
P(\xi_1,\xi_2)
d\xi_1 d\xi_2
}
\nonumber \\
&=&
\frac{
\int_0^{2\pi} \int_0^{2\pi}
P(x_1|\alpha,\xi_1)P(x_2|\beta,\xi_2)
w(\alpha,\beta,\xi_1,\xi_2,W)
P(\xi_1,\xi_2)
d\xi_1 d\xi_2
}{
\int_0^{2\pi} \int_0^{2\pi}
w(\alpha,\beta,\xi_1,\xi_2,W)
P(\xi_1,\xi_2)
d\xi_1 d\xi_2
}
,
\end{eqnarray}
where, in general, the weight function
\begin{eqnarray}
\label{pepr3}
w(\alpha,\beta,\xi_1,\xi_2,W)
&=&
\int_{-\infty}^{+\infty}dt_1 \int_{-\infty}^{+\infty} dt_2\,
P(t_1|\alpha,\xi_1)
P(t_2|\beta,\xi_2)
\Theta(W-|t_1 -t_2|)
,
\end{eqnarray}
will be less than one (because
$\int_{-\infty}^{+\infty}dt_1 \int_{-\infty}^{+\infty} dt_2\,P(t_1|\alpha,\xi_1)P(t_2|\beta,\xi_2)=1$)
unless $W$ is larger than the range of $(t_1,t_2)$ for which $P(t_1|\alpha,\xi_1)$ and
$P(t_2|\beta,\xi_2)$ are nonzero.
It is self-evident that unless $w(\alpha,\beta,\xi_1,\xi_2,W)=w(\alpha,\xi_1,W)w(\beta,\xi_2,W)$,
 Eq.~(\ref{pepr2}) cannot be written in the factorized form
$P(x_1,x_2|\alpha,\beta)=\int P(x_1|\alpha,\lambda)P(x_2|\beta,\lambda)\rho(\lambda)d\lambda$
(see Ref.~\onlinecite{BELL93,BALL03} for the notation) that is essential
to derive the original Bell (CHSH) inequalities.

In our simulation model, the time delays $t_i$ are distributed uniformly
over the interval $[0,T_i]$ where $T_1=T_0|\sin 2(\alpha-\xi_1)|^d$ and
$T_2=T_0|\sin 2(\beta-\xi_2)|^d$.
Thus, $P(t_1|\alpha,\xi_1)= \Theta(t_1)\Theta(T_1 - t_1)/T_1$,
$P(t_2|\beta,\xi_2)= \Theta(t_2)\Theta(T_2 - t_2)/T_2$, and
\begin{eqnarray}
\label{pepr4}
w(\alpha,\beta,\xi_1,\xi_2,W)
&=&\frac{1}{T_1T_2}
\int_0^{T_1}dt_1 \int_0^{T_2} dt_2\,
\Theta(W-|t_1 -t_2|)
.
\end{eqnarray}
The integrals in Eq.(\ref{pepr4}) can be worked out analytically, yielding
\begin{eqnarray}
\label{pepr5}
w(\alpha,\beta,\xi_1,\xi_2,W)
=\frac{1}{4T_1T_2}&[&T_1^2+T_2^2+2(T_1+T_2)W+(W-T_1)|W-T_1|
\nonumber \\
&&+(W-T_2)|W-T_2|-(W-T_1+T_2)|W-T_1+T_2|
\nonumber \\
&&-(W+T_1-T_2)|W+T_1-T_2|\;\;]
.
\end{eqnarray}
Clearly, Eq.~(\ref{pepr5}) cannot be written in the factorized form $w(\alpha,\xi_1,W)w(\beta,\xi_2,W)$.
Hence, it should not come as a surprise that as soon as we want to simulate the real EPRB experiment
with photons in which the time window is essential,
we can obtain correlations that cannot be described by Bell-like models.

According to our simulation model (and the assumption made at the beginning of this section),
the probability distributions that describe the polarizers are given by
\begin{eqnarray}
\label{pepr6}
P(x_1|\alpha,\xi_1)
&=&\frac{1+x_1\cos2(\alpha-\xi_1)}{2},
\nonumber \\
P(x_2|\beta,\xi_2)&=&\frac{1+x_2\cos2(\beta-\xi_2)}{2}
.
\end{eqnarray}
It is easy to check that these distributions reproduce Malus law for a single polarizer.

We now consider some specific cases.
First, we consider Case I and specialize to the case
that the source emits particles with opposite polarization
$P(\xi_1,\xi_2)=\delta(\xi_1+\pi/2-\xi_2)P(\xi_1)$
with $P(\xi_1)$ being a uniform distribution.
If $d=0$ and $W\le T_0$, we have $w(\alpha,\beta,\xi_1,\xi_2,W)=(2T_0-W)W/T_0^2$.
Likewise, if $W>T_0$, $w(\alpha,\beta,\xi_1,\xi_2,W)=1$.
Therefore, if $W>T_0$ or $d=0$, we have
\begin{eqnarray}
\label{pepr11}
P(x_1,x_2|\alpha,\beta)
&=&
\frac{
\int_0^{2\pi} \int_0^{2\pi}
P(x_1|\alpha,\xi_1)P(x_2|\beta,\xi_2)
P(\xi_1,\xi_2)
d\xi_1 d\xi_2
}{
\int_0^{2\pi} \int_0^{2\pi}
P(\xi_1,\xi_2)
d\xi_1 d\xi_2
}
\nonumber \\
&=&
\int_0^{2\pi} \int_0^{2\pi}
P(x_1|\alpha,\xi_1)P(x_2|\beta,\xi_2)
P(\xi_1,\xi_2)
d\xi_1 d\xi_2
\nonumber \\
&=&
\frac{1}{8\pi}
\int_0^{2\pi}
(1+x_1\cos2(\alpha-\xi))(1-x_2\cos2(\beta-\xi))
d\xi
\nonumber \\
&=&
\frac{2+x_1x_2\cos2(\alpha-\beta)}{8}
,
\end{eqnarray}
showing that if we ignore the time-tag information, the two-particle probability
takes the form of the hidden variable models considered by Bell~\cite{BELL93},
and we cannot reproduce the results of quantum theory~\cite{BELL93}.

Second, we consider Case I but focus on the regime for small $W$,
the regime that experimenters aim to reach~\cite{WEIH00}.
Then, Eq.~(\ref{pepr5}) reduces to
\begin{eqnarray}
\label{pepr7}
w(\alpha,\beta,\xi_1,\xi_2)
=\frac{2W}{\max(T_1,T_2)}+{\cal O}(W^2)
.
\end{eqnarray}
and we find that $E_1(\alpha,\beta)=E_2(\alpha,\beta)=0$ and that
\begin{eqnarray}
\label{pepr9}
E(\alpha,\beta)
&=&
-
\frac{
\int_0^{2\pi} \cos2(\xi-\alpha)\cos2(\xi-\beta)\max(|\sin 2(\xi-\alpha)|,|\sin 2(\xi-\beta)|)^{-d} d\xi
}{
\int_0^{2\pi} \max(|\sin 2(\xi-\alpha)|,|\sin 2(\xi-\beta)|)^{-d} d\xi
}
\nonumber \\
&=&
-
\frac{
\int_{\theta/2}^{\theta/2+\pi/4} \cos2\xi\cos2(\xi-\theta)|\sin 2\xi|^{-d} d\xi
}{
\int_{\theta/2}^{\theta/2+\pi/4} |\sin 2\xi|^{-d} d\xi
}
\nonumber \\
&=&
\left\{
\begin{array}{ll}
-\frac{1}{2}\cos 2\theta&, d=0\\
\frac{\pi}{4}\sin2\theta\cos2\theta - \cos2\theta +\ln[|\tan \theta|^{\sin^22\theta/2}]&, d=2\\
-\cos 2\theta&, d=4\\
-\frac{1}{2}\cos 2\theta\left[1+24(19+5\cos4\theta)^{-1}\right]&, d=6\\
-(53\cos2\theta+7\cos6\theta)(39+21\cos4\theta)^{-1}&, d=8
\end{array}
\right.
,
\end{eqnarray}
where $\theta=\alpha-\beta$ and we have omitted the expressions for odd $d$ because they cannot be written
in terms of elementary functions.
Note that by passing to the continuum limit, one has to be careful with the integrals that
appear in expressions such as Eq.~(\ref{pepr9}): The ratio of the two integrals is well-defined
but each individual integral may vanish or diverge for some choices of $\alpha-\beta$.
Needless to say, such situations do not occur in the (discrete) simulation model.

Finally, in Case II, the values of $\xi_1$ and $\xi_2$ are fixed,
hence $P(\xi_1,\xi_2)=\delta(\xi_1-\eta_1)\delta(\xi_2-\eta_2)$.
Then, as is clear from Eq.~(\ref{pepr2}), the weight function $w(\alpha,\beta,\eta_1,\eta_2,W)$
drops out and the two-particle probability reduces to
\begin{eqnarray}
\label{pepr12}
P(x_1,x_2|\alpha,\beta)
&=&
P(x_1|\alpha,\eta_1)P(x_2|\beta,\eta_2)
,
\end{eqnarray}
such that
\begin{eqnarray}
\label{pepr18}
E(\alpha,\beta,W)&=&\sum_{x_1,x_2=\pm1}x_1x_2P(x_1,x_2|\alpha,\beta)=\left(\sum_{x_1=\pm1}x_1P(x_1|\alpha,\beta)\right)
\left(\sum_{x_2=\pm1}x_2P(x_2|\alpha,\beta)\right)
\nonumber \\
&=&
E_1(\alpha,\beta,W)E_2(\alpha,\beta,W)
.
\end{eqnarray}
Evidently, the simulation model will reproduce the results of quantum theory
for Case II if the proper expression, the one yielding Malus law, is used
for the single-particle probabilities $P(x_1|\alpha,\eta_1)$ and $P(x_2|\beta,\eta_2)$.

Summarizing: Up to first order in the time window $W$ and for $d=4$,
in Case I (corresponding to the case in which the source emits particles with opposite random polarization)
the probabilistic model of the simulation algorithm yields
\begin{eqnarray}
\label{pepr19}
E_1(\alpha,\beta)&=&E_2(\alpha,\beta)=0
\nonumber \\
E(\alpha,\beta)&=&-\cos2(\alpha-\beta)
,
\end{eqnarray}
for the single-particle averages and two-particle correlation, respectively.
Obviously, these expressions are identical to those given in the second
column of Table~\ref{tab1}.
If, as in Case II, the source emits particles with fixed polarizations $\eta_1$ and $\eta_2$,
respectively, the probabilistic model of the simulation algorithm yields
\begin{eqnarray}
\label{pepr20}
E_1(\alpha,\beta)&=&\cos2(\alpha-\eta_1)\quad,\quad E_2(\alpha,\beta)=\cos2(\alpha-\eta_2)
\nonumber \\
E(\alpha,\beta)&=&-\cos2(\alpha-\eta_1)\cos2(\alpha-\eta_2)
,
\end{eqnarray}
in exact agreement with the results in the third column of Table~\ref{tab1}.
Thus, it follows that to first order in $W$, the probabilistic model of the simulation algorithm can reproduce {\sl exactly}
the results for the single- and two-particle averages of
the quantum theory of a system of two photon polarizations.

\end{widetext}

\section{Simulation results}\label{sec5}
\label{sec:results}

We use the computer model, described earlier to simulate Cases I and II.
The simulation proceeds in exactly the same way as in the experiment, that is we first collect
the data sets $\Upsilon_1$ and $\Upsilon_2$ for various settings of the polarizers
(various $\theta_{n,i}$), and then compute the coincidences Eq.(\ref{Cxy}), the
average single-particle counts  and the correlation Eq.(\ref{Exy}),
from which we can calculate the function $S(\theta)$
(see Eqs.~(\ref{eq29}) and (\ref{eq31})).
The parameters for all simulations are $k=1$, $d=4$, $\tau=0.00025$, and $N=10^6$,
unless mentioned otherwise.

\begin{figure*}[t] 
\begin{center}
\mbox{
\includegraphics[width=8cm]{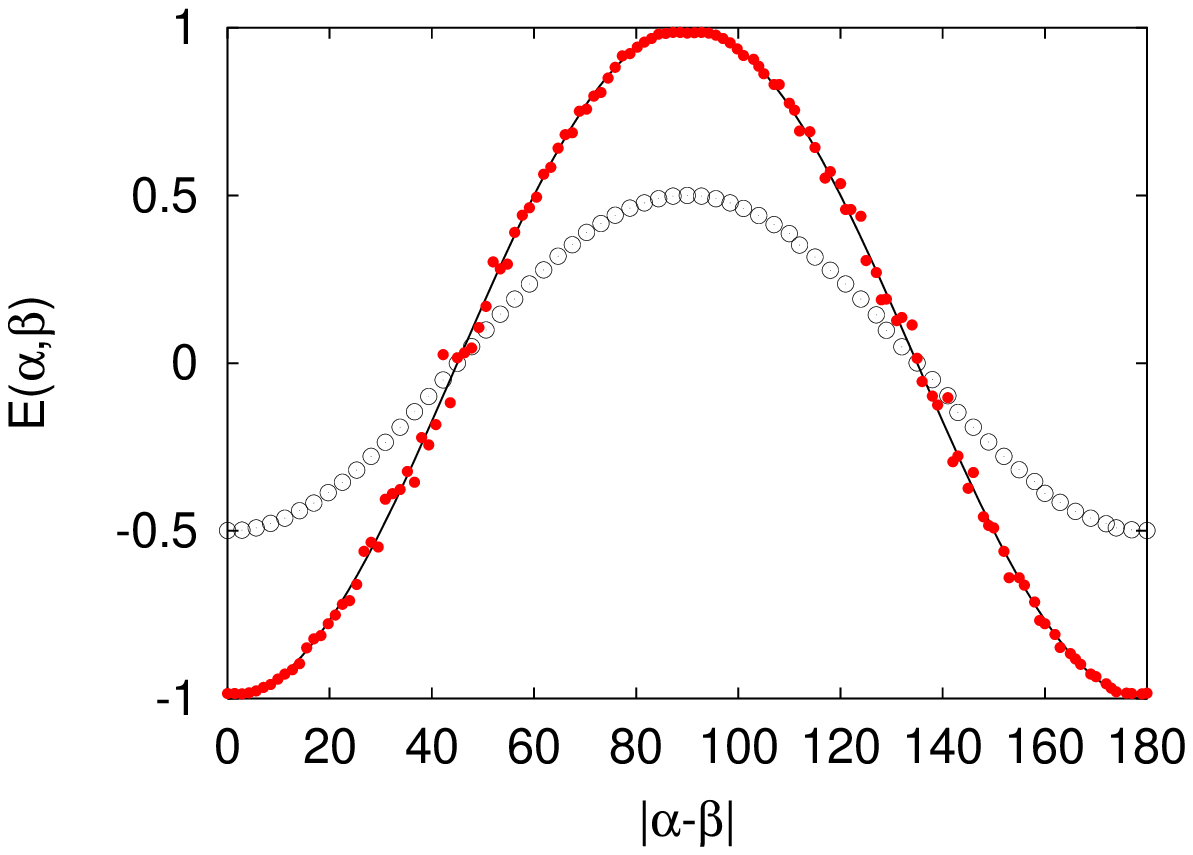}
\includegraphics[width=8cm]{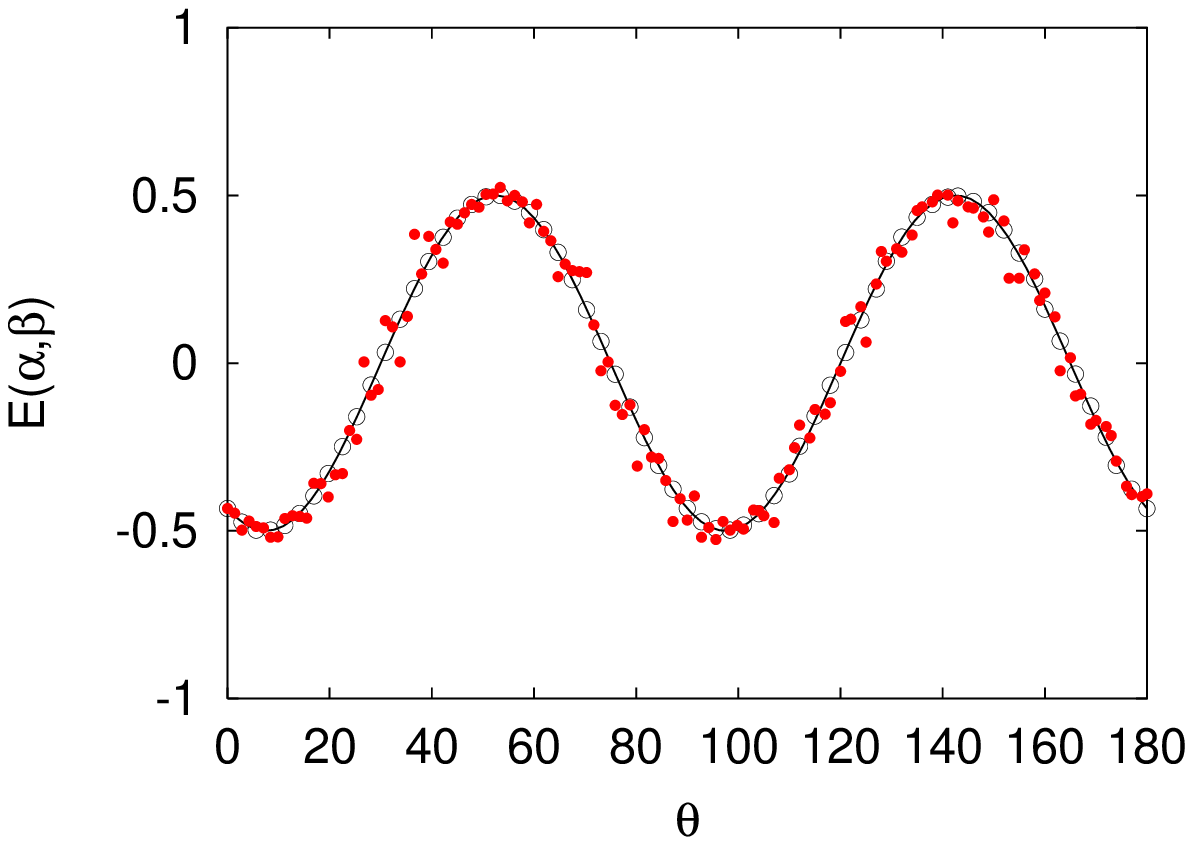}
}
\caption{(color online)
Correlation $E(\alpha ,\beta )$ between the coincidence counts as a function of the
orientation difference of the two polarizers in each observation station.
Left: Computer simulation of Case I in which the
source emits particles with opposite random polarization (EPRB experiment).
Right: Computer simulation of Case II in which the source emits particles with fixed polarization and
 $\alpha =\theta $, $\beta=\theta +\pi /4$, $\eta_1=\pi /6$, and $\eta_2=\pi /6+\pi /2$ (see Fig.~\ref{fig1b}).
Squares (red): Simulation results using the time-delay mechanism (with $d=4)$ to compute the two-particle coincidence.
Open circles (black): Simulation results without using the time-tags (equivalent to $d=0$ or $W\rightarrow\infty$).
Solid lines: Quantum theory.
}
\label{fig10}
\end{center}
\end{figure*}

In Fig.~\ref{fig10} (left), we present simulation data for the correlation $E(\alpha,\beta)$ for Case I, that is
for the case that the source emits particles with an opposite, random
polarization, corresponding to the singlet state in the quantum theoretical
description.
Figure~\ref{fig10} (right), shows the corresponding results for Case II.
It is clear that in both cases, the agreement between the simulation data and
quantum theory is excellent.

Also shown in Fig.~\ref{fig10} are the results for $E(\alpha,\beta)$ if we ignore the time-delay data
(equivalent to $d=0$ or $W\rightarrow\infty$). In Case I we obtain simulation results
that agree very well with the expression $E(\alpha,\beta )=-(1/2)\cos 2\theta$ (see Eq.(\ref{pepr9})),
a result that differs from what is obtained by considering the class of models studied by Bell~\cite{BELL93}.
In the latter case an equilateral saw-tooth function is obtained instead of the cosine.
In Case II, the results for $d=4$ and $d=0$ or $W\rightarrow\infty$ are, apart from statistical fluctuations, the same.
Hence, for Case II the time window $W$ can be omitted for the calculation of the two-particle
correlation function.

\begin{figure*}[t] 
\begin{center}
\mbox{
\includegraphics[width=8cm]{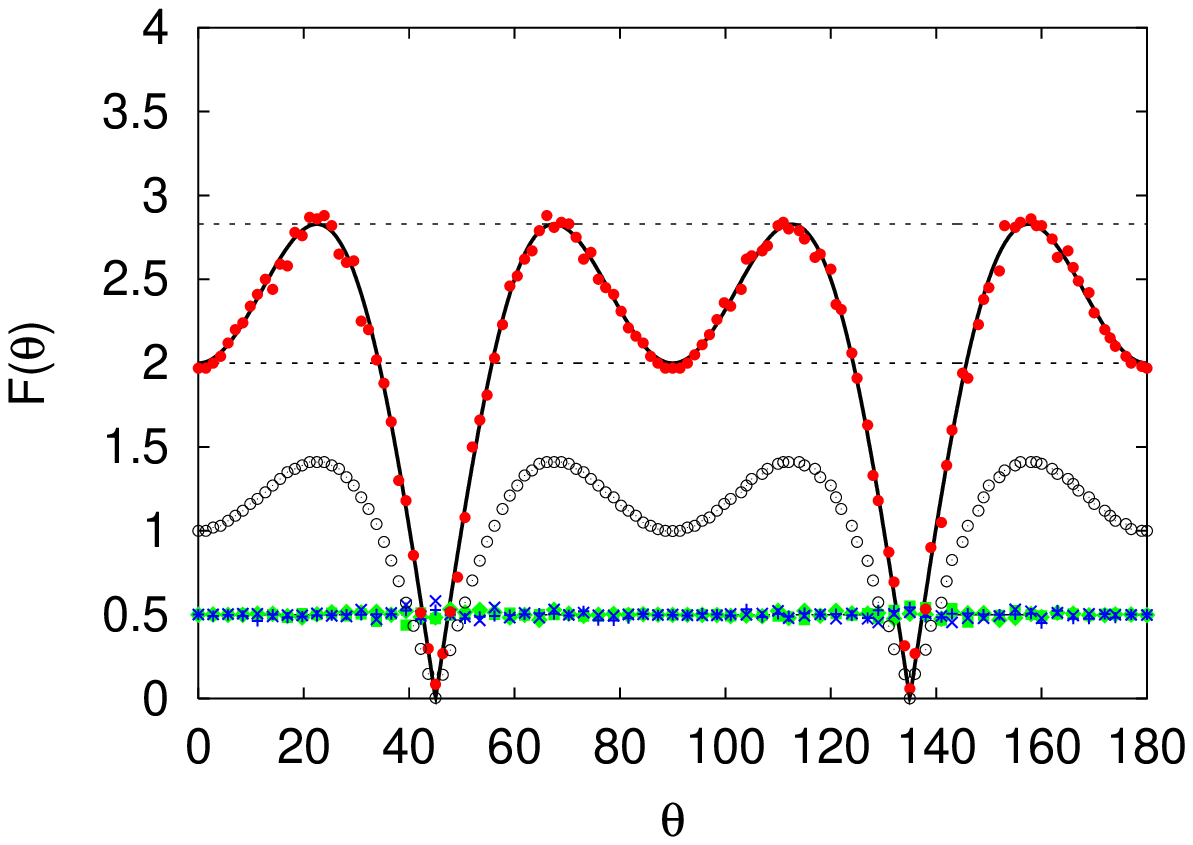}
\includegraphics[width=8cm]{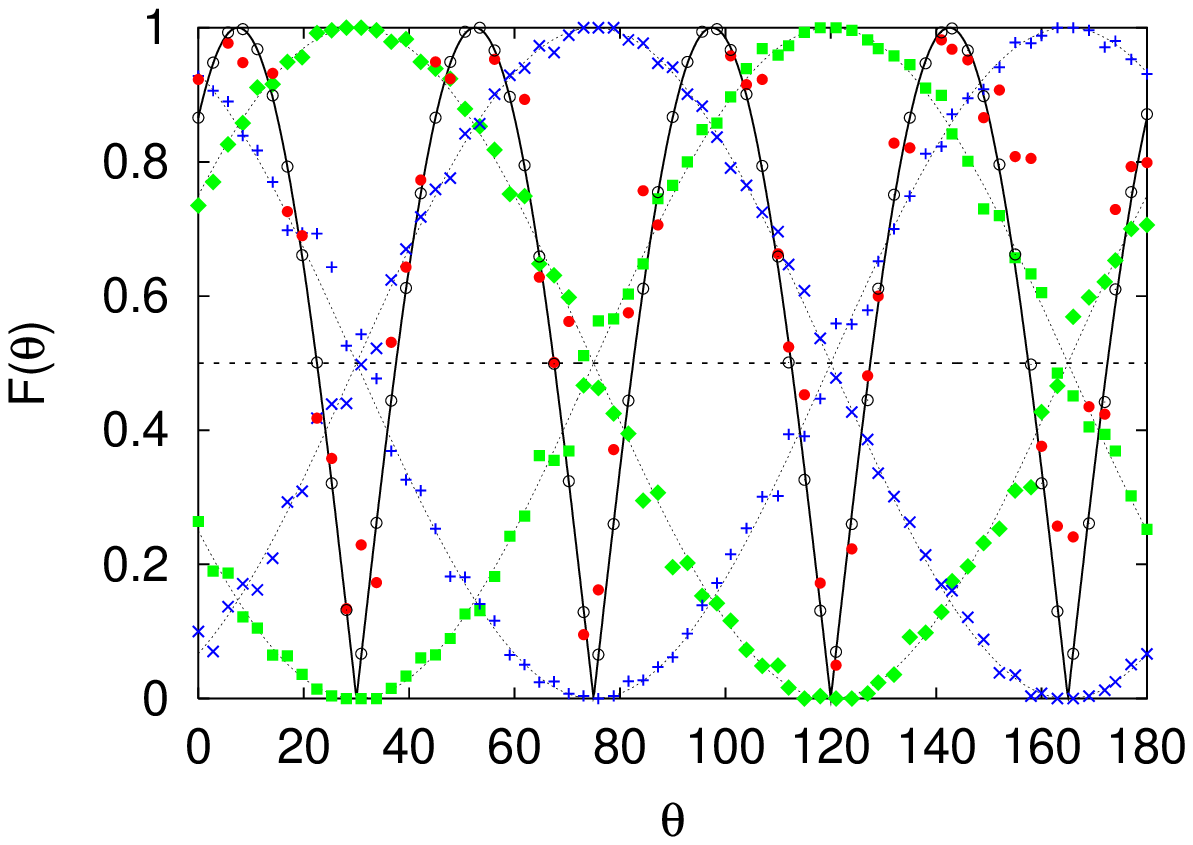}
}
\caption{(color online)
Left: Computer simulation of Case I in which the
source emits particles with opposite random polarization (EPRB experiment).
Right: Computer simulation of Case II in which the source emits particles with fixed polarization.
Squares (red): Simulation results for $F(\theta)=|S(\theta)|$ using the time-delay mechanism ($d=4)$.
Open circles (black): Simulation results for $F(\theta)=|S(\theta)|$ without using the time-tags (equivalent to
$d=0$ or $W\rightarrow\infty$).
Other markers: Average single-particle counts on the detectors (see Fig.~\ref{aspect}).
Squares (green): $F(\theta=\theta_1)=P_{+}(\theta_1)$;
Diamonds (green): $F(\theta=\theta_1)=P_{-}(\theta_1)$;
Plusses (blue): $F(\theta=\theta_2)=P_{+}(\theta_2)$;
Crosses (blue): $F(\theta=\theta_2)=P_{-}(\theta_2)$.
In Case I (left), these four symbols lie on top of each other.
In Case II (right), these markers show the typical Malus law behavior.
Solid line: Quantum theory for $|S(\theta)|$.
Dashed line at $|S(\theta )|=2\sqrt 2 $: Maximum of $S(\theta )$ if the system is described by quantum theory.
Dashed line at $|S(\theta )|=2$: Maximum of $S(\theta )$ if the system is described by the
class of models introduced by Bell~\cite{BELL93};
Dashed line at $|S(\theta)|=1/2$: Expected number of $+1$ and $-1$ events recorded by the detectors
if the input to the polarizers consists of particles with random polarization.
Dotted lines: Quantum theory for $P_{+}(\theta_1)$, $P_{-}(\theta_1)$, $P_{+}(\theta_2)$ and $P_{-}(\theta_2)$.
}
\label{fig11}
\end{center}
\end{figure*}

\begin{figure*}[t] 
\begin{center}
\mbox{
\includegraphics[width=8cm]{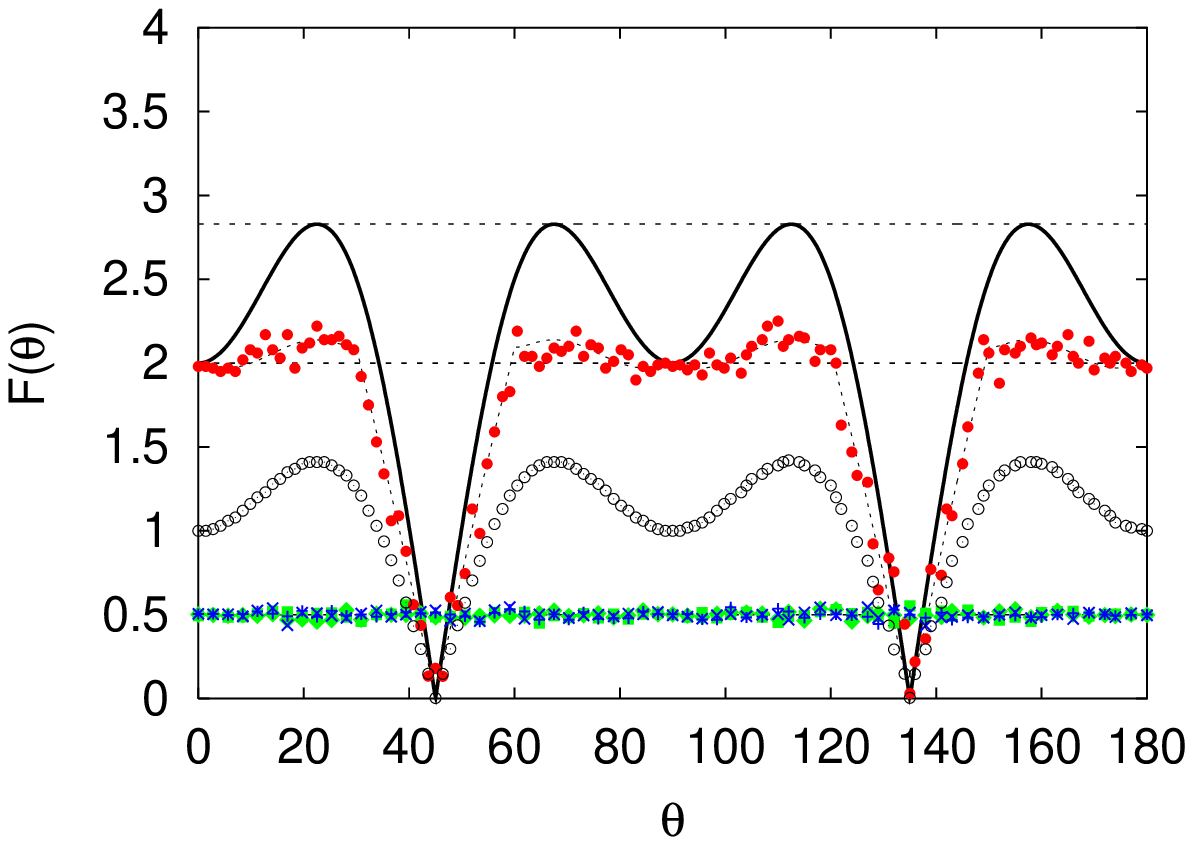}
\includegraphics[width=8cm]{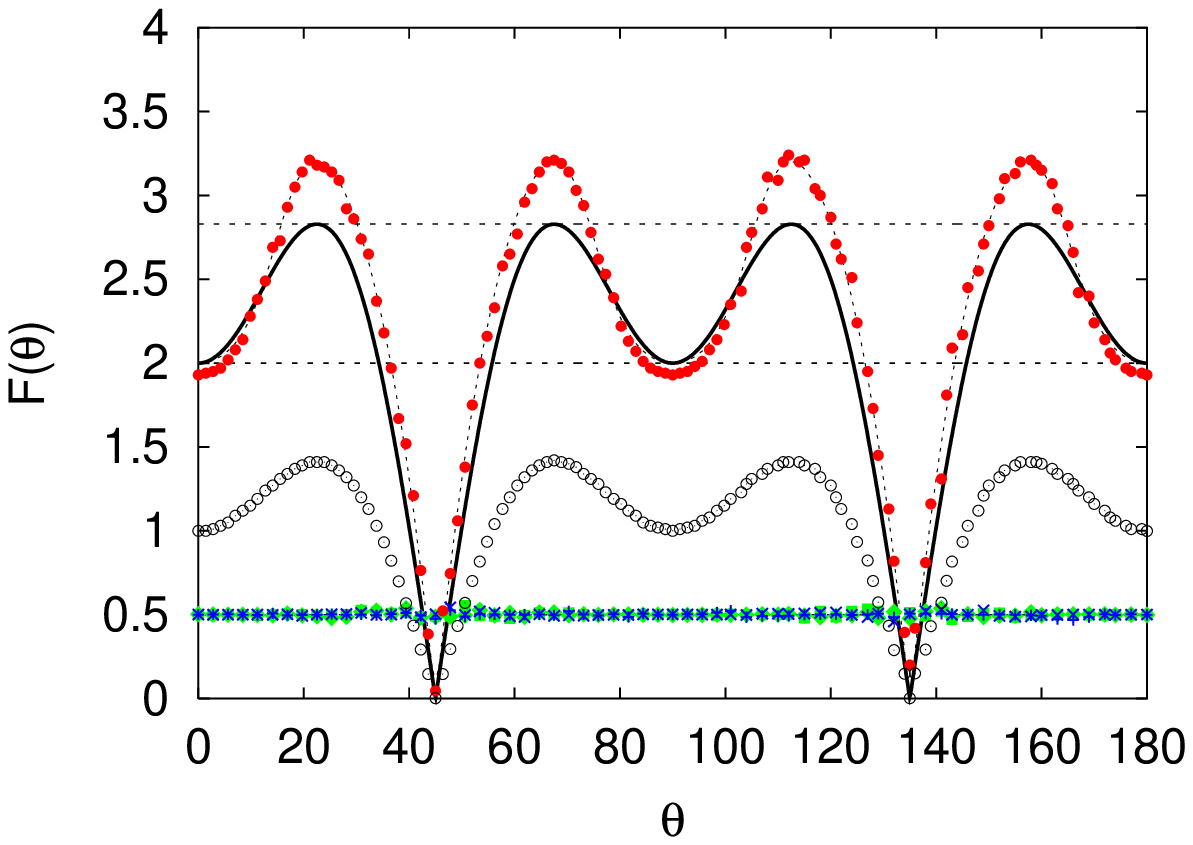}
}
\caption{(color online)
Left: Same as Fig.~\ref{fig11} (left) except that $d=2$.
Dotted line: $F(\theta)=|S(\theta)|$ calculated from Eq.(\ref{pepr9}).
Right: Same as Fig.~\ref{fig11} (left) except that $d=6$.
Dotted line: $F(\theta)=|S(\theta)|$ calculated from Eq.(\ref{pepr9}).
}
\label{fig12}
\end{center}
\end{figure*}

In Fig.~\ref{fig11} (left), we present additional simulation data for Case I.
It is clear that for $d=4$, the simulation model reproduces the results
of quantum theory for the single-particle expectation values $P_{\pm}(\alpha)$ and $P_{\pm}(\beta)$ (see Table~\ref{tab1})
and $S(\theta)$ (see Eq.~(\ref{Stheta})).
Indeed, the frequency with which each detector fires is approximately one-half and
$|S(\theta)|$ agrees with the expressions $E(\alpha,\beta)=-\cos 2(\alpha-\beta)$
that is obtained for the singlet state.
Also shown in Fig.~\ref{fig11} (left) are the results for $|S(\theta)|$
if we ignore the time-tag data.
Effectively, this is the same as letting the time window $W\rightarrow\infty$ or
setting $d=0$. Then, our simulation model generates data that
satisfies $|S(\theta)|\le2$, which is what we expect for the
class of models studied by Bell~\cite{BELL93}.

In Case II, the source emits particles with a fixed (but not necessarily opposite)
polarization. In the right panel of Fig.~\ref{fig11}, we present results
for the case $\theta_1=\alpha =\theta $ and
$\theta_2=\beta=\theta +\pi /4$.
The angle $\xi $ of the particles is $\pi /6$
(corresponding to $\eta_1=\pi /6$ and $\eta_2=\pi /6+\pi /2$ in the quantum theoretical description).
For this choice, we have $P_+ (\alpha)=\cos ^2(\theta -\pi/6 )$,
$P_+ (\beta)=\cos ^2(\theta -\pi /6 -\pi /4)$, $E(\alpha,\beta)=2^{-1}\sin
4(\pi /6-\theta)$ and $S(\theta )=\sin 4(\pi /6-\theta)$.
Also seen from Fig.~\ref{fig11} (right) is that $|S(\theta)|$ does not depend on $d$,
that is apart from statistical fluctuations, the time-tag data do not affect $|S(\theta)|$.

From Fig.~\ref{fig11}, it is clear that for $d=4$, the event-by-event simulation model
reproduces the single- and two-particle results of quantum theory for both
Case I and II, {\sl without any change to the algorithm that simulates the
polarizers}.

Having established that the data generated by our ``non-quantum'' system
agrees with quantum theory, it is of interest to explore if
these algorithms can generate data that is not described by
quantum theory and by the locally causal, probabilistic models introduced by Bell~\cite{BELL93}.
We can readily give an affirmative answer to this question by
repeating the simulations for Case I (see Fig.~\ref{fig11} (left)) for different
values of the time-delay parameter $d$, all other parameters being the same as
those used to obtain the data presented in Fig.~\ref{fig11}.

For $d=0$, simulations with or without time-delay mechanism yield data
that, within the usual statistical fluctuations, are the same (results not shown)
and satisfy $|S(\theta)|\le2$ .
Figure~\ref{fig12} shows the simulation data for $d=2$ and $d=6$. For $0<d<4$ our model yields
two-particle correlations that are stronger than those of the Bell-type
models but they are weaker than in the case of the singlet state in quantum theory.
Therefore, the maximum of $S(\theta )$ is less than $2\sqrt 2$ but larger than two.
For $d\ge5$, we find that the two-particle correlations are
significantly stronger than in the case of the singlet state in quantum theory.

For $d<4$, $2\le S_{max}<2\sqrt{2}$ for any value of $W/\tau$. Hence, for $d<4$ our model cannot produce the
correlations of the singlet state.
For $d=4$, $2\le S_{max}\le 2\sqrt{2}$ and our model produces the correlations of the singlet state if $W$
is sufficiently small such that contributions of order $W^2$ can be negelected.
For $d>4$, $2\le S_{max}\le 4$, and for a range of $W/\tau$, $S_{max}>2\sqrt{2}$, implying that our model
exhibits correlations that cannot be described by the quantum theory of two spin-1/2 particles, while
still satisfying Einstein's criteria for local causality.

From Fig.~\ref{fig12} it can be seen that for $d=2$ and $d=6$ there is
good agreement between the results obtained with our event-based simulation model and
the analytical result for $|S(\theta )|$ obtained from Eq.~(\ref{pepr9}).
For $d=2$, the simulation results show larger fluctuations than for $d=6$, but in all cases these fluctuations
can be reduced by increasing $N$ (results not shown).

\begin{figure*}[t]
\begin{center}
\mbox{
\includegraphics[width=8cm]{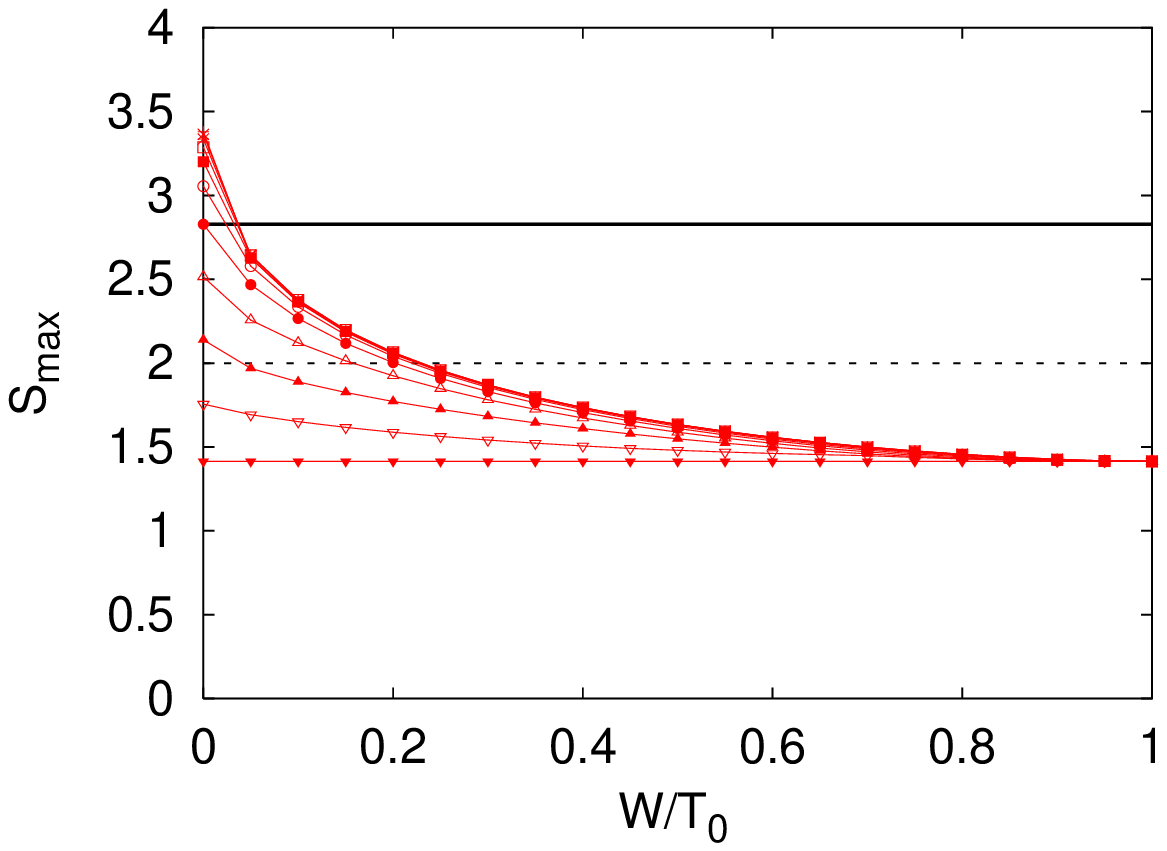}
\includegraphics[width=8cm]{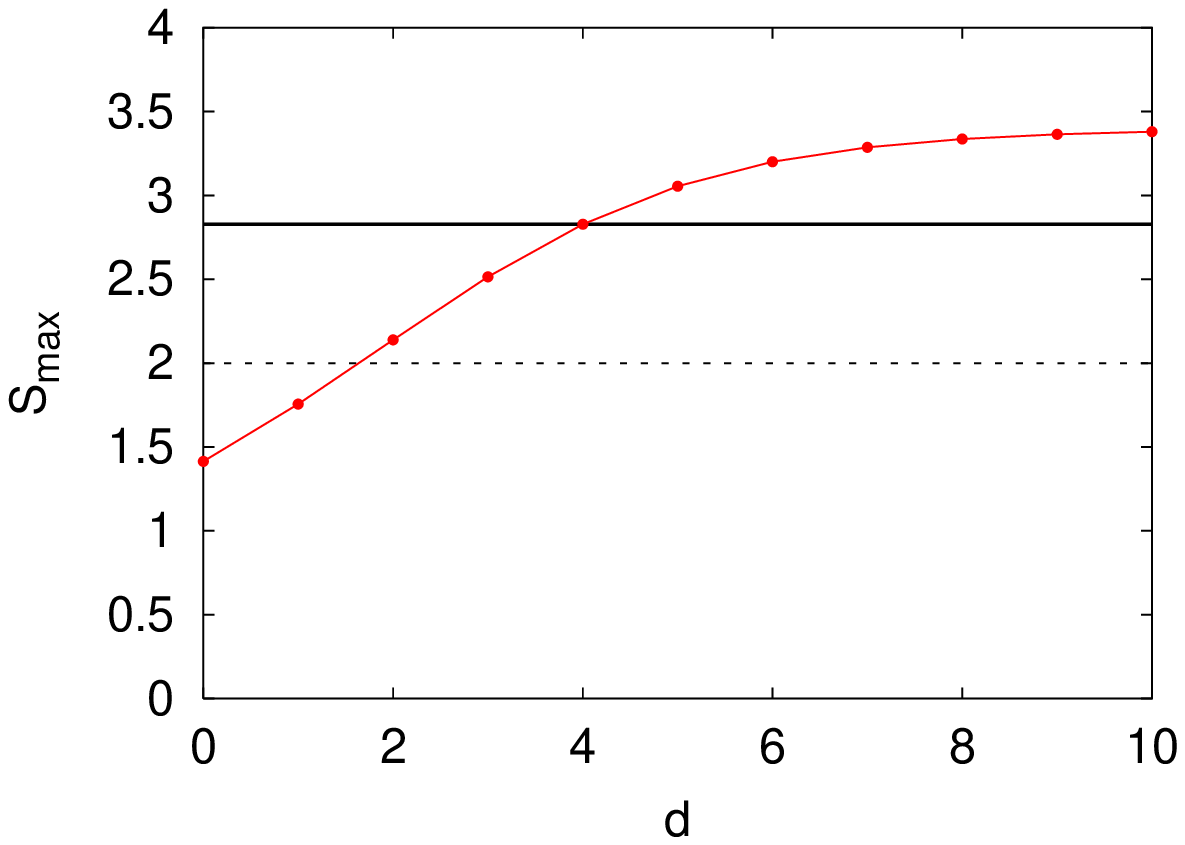}
}
\caption{(color online)
Left: $S_{max}=\max_{\alpha,\alpha',\beta,\beta'} S(\alpha,\alpha',\beta,\beta')$
as a function of the time window $W$ relative to the maximum time delay resolution $T_0$.
Curves from bottom to top: Results for $d=0, 1, \ldots , 10$.
Solid line at $2\sqrt{2}$: Value of $S_{max}$ for a quantum system in the singlet state.
Dashed line at $2$: Value of $S_{max}$ for a quantum system in an uncorrelated state.
Right: $S_{max}=\max_{\alpha,\alpha',\beta,\beta'} S(\alpha,\alpha',\beta,\beta')$
as a function of the time-delay parameter $d$ negelecting contributions of ${\cal O}(W)$.
Solid line at $2\sqrt{2}$: Value of $S_{max}$ for a quantum system in the singlet state.
Dashed line at $2$: Value of $S_{max}$ for a quantum system in an uncorrelated state.
}
\label{maxStheta}
\end{center}
\end{figure*}

The simulation results presented in Figs.~\ref{fig11} and \ref{fig12}
have been obtained for $W/\tau=1$ and small $\tau$ (recall that the unit of time
in our numerical work is set equal to one).
In general, in experiment the two-particle correlation depends on both $W$ and $\tau$.
Our simulation model makes definite predictions for this dependence.
This can be seen from Fig.~\ref{maxStheta}, showing
$S_{max}=\max_{\alpha,\alpha',\beta,\beta'} S(\alpha,\alpha',\beta,\beta')$
as a function of $W/T_0$ for various values of $d$
and of $S_{max}$ to first order in $W$ as a function of $d$.
The numerical results agree with the values of $S_{max}$ that have been obtained analytically
to first order in $W$, $d=0,2,...,8$ and for $W\rightarrow\infty$.

Summarizing: In the regime of small $W/T_0=\tau/T_0$, the results produced by the simulation
algorithm are in excellent agreement with the quantum theoretical expressions (see Table~\ref{tab1})
of the single- and two-particle averages.

\section{Comparison with experimental data}\label{sec5a}

For the simulation model described in Section~\ref{sec4},
it follows immediately from Eqs.~(\ref{Cxy}) and (\ref{Exy}) that $|E(\alpha,\beta)|<1$ and that
\begin{equation}
\label{eq400}
|E(\alpha ,\beta )-E(\alpha,{\beta }')+E({\alpha }',\beta )+E({\alpha }',{\beta }')|\le4.
\end{equation}
Without any further constraints on the algorithm that generates the
data $\{\Upsilon_1,\Upsilon_2\}$ 
the upperbound (4) in Eq.~(\ref{eq400}) cannot be improved.
On the other hand, for a local realist (probabilistic) model, it can be shown that~\cite{LARS04}
\begin{equation}
\label{eq401}
|E(\alpha ,\beta )-E(\alpha,{\beta }')+E({\alpha }',\beta )+E({\alpha }',{\beta }')|\le\frac{6}{\gamma}-4,
\end{equation}
where $\gamma$ is the infimum of the probability of coincidence over all possible settings $\{\alpha,\beta\}$.
In our simulation model the frequency of coincidences
\begin{equation}
\label{eq402}
\Gamma=\frac{1}{N}\sum_{n=1}^{N}\Theta(W-\vert t_{n,1} -t_{n ,2}\vert)
,
\end{equation}
is easy to compute and, assuming that the results that we obtain by using pseudo-random numbers
can be described by a probabilistic model (see Section~\ref{sec7}), we may assume that $\gamma=\Gamma$
with probability one. For $W=\tau$ and $d=4$, a straightforward calculation gives
\begin{equation}
\label{eq402a}
\gamma=\min_{\alpha-\beta}\Gamma =\frac{16}{3\pi} \frac{W}{T_0}+ {\cal O}(W^2)
,
\end{equation}
showing that up to first order in the time window $W$, the minimum frequency of coincidences
is proportial to the time window, as one naively would expect.

As pointed out in Ref.~\cite{LARS04}, a local realist model
that uses coincidence in time to decide which particles form a pair is not necessarily in conflict
with the predictions of quantum theory unless $\gamma>3-3/\sqrt{2}$.
Thus, in Case I, it is of interest to explore how $\Gamma$ affects $S_{max}$ and the sinusoidal
shape of the two-particle correlation but before we present some results, we want to draw attention to the fact that
the model that we introduce in this paper is not unique in the sense that it is not
the only model that reproduces the results of quantum theory for the singlet state~\cite{RAED06c}.
Different models will yield different numerical results for $\Gamma$ but the general behavior
is the same.
In these event-based models, there are three independent parameters that we can use to ``tune'' the
simulation results to experimental data, namely $\tau/T_0$, $W/T_0$, and $d$.

First, we consider the problem of ``fitting'' our model to experimental data for $S_{max}$.
As explained earlier, to obtain $S_{max}$, one has to perform
four experiments.
For instance, Ref.~\cite{WEIH98} reports $S_{max}\approx 2.73$
using  $\alpha=\theta_1=0,\pi/4$ and $\beta=\theta_2=\pi/8,3\pi/8$
and if we make the hypothesis that the frequency of coincidences
that we found earlier ($\approx0.01$ for $\alpha-\beta=\pm\pi/8,-3\pi/8$) is a good estimate for $\gamma$,
no conclusion can be drawn from the relevant theoretical bound Eq.~(\ref{eq401}),
other than that this experiment does not rule out a local realist (probabilistic) description~\cite{LARS04}.
The simulation model described in Section~\ref{sec4},
reproduces the experimental result $S_{max}\approx2.73$~\cite{WEIH98} for
$\tau/T_0=1/29$, $W/T_0=1/29$, and $d=4$.
For $\alpha-\beta=0,\pi/8,\pi/4,3\pi/8,\pi/2$ we find $\Gamma=0.38,0.14,0.06,0.14,0.38$, respectively.
Because in our simulation the source only emits particle pairs and since no particles are lost or falsely detected,
we may expect that the simulation for $\alpha-\beta=\pm\pi/8,-3\pi/8$
yields a value of $\Gamma$ that is larger than the one ($\approx0.01$) extracted from
the experimental data~\cite{WEIH98}.
A simulation run with N=300000 events
(roughly the same number as observed in the experimental data analyzed in Section~\ref{sec2a}),
gives $\sum_n x_{n,1}= 0.0016,-0.0011$ for $\alpha=0,\pi/4$
and $\sum_n x_{n,2}=0.007,0.001$ for $\beta=\pi/8,3\pi/4$, respectively,
in reasonable agreement with the experimental results (see Section~\ref{sec2a}).

For comparison, the simulation model introduced in Ref.~\cite{RAED06c}
reproduces the same value of $S_{max}$ for
$\tau/T_0=1/9$, $W/T_0=1/9$, and $d=2$,
yielding $\Gamma=0.38,0.18,0.13,0.18,0.38$
for $\alpha-\beta=0,\pi/8,\pi/4,3\pi/8,\pi/2$, respectively.

As another example, we consider the result $S_{max}\approx2.25$
as obtained from ion-trap experiments~\cite{ROWE01}.
Although it is not evident that the events registered in this experiment
are as simple as the detection of single photons, let us assume
that the model for the real EPRB experiment with photons can
nevertheless be used to describe the outcome of these ion-trap experiments.
Then, the simulation model described in Section~\ref{sec4} reproduces the value of $S_{max}\approx2.25$ if we take
$\tau/T_0=1/4.3$, $W/T_0=1/4.3$, and $d=4$.
For $\alpha-\beta=0,\pi/8,\pi/4,3\pi/8,\pi/2$ we find $\Gamma=0.65,0.46,0.32,0.46,0.65$, respectively.
For comparison, the simulation model described in Ref.~\cite{RAED06c}
yields the same value of $S_{max}=2.25$ for
$\tau/T_0=1/1.031$, $W/T_0=1/1.031$, and $d=2$,
with $\Gamma=0.95,0.89,0.89,0.89,0.95$
for $\alpha-\beta=0,\pi/8,\pi/4,3\pi/8,\pi/2$, respectively.
As $\Gamma\approx1$, this experiment seems to have an almost ideal detection efficiency~\cite{ROWE01}.

For completeness, we consider the case $S_{max}=2.83$.
Recall that both the model introduced in this paper and the one of Ref.~\cite{RAED06c} reproduce the result
($E(\alpha-\beta)=-\cos2(\alpha-\beta)$)
of quantum theory if we keep the contributions of ${\cal O}(W)$ only.
The model described in Section~\ref{sec4} yields $S_{max}=2.83$ if we take
$\tau/T_0=1/1500$, $W/T_0=1/1500$, and $d=4$
for which
$\Gamma=0.13,0.0032,0.0011,0.0032,0.13$
for
$\alpha-\beta=0,\pi/8,\pi/4,3\pi/8,\pi/2$, respectively.
For these choices of parameters, the numerical results for $E(\alpha-\beta)$ are very close to
those of quantum theory (see Fig.~\ref{fig10}).

For the model of Ref.~\cite{RAED06c} and $\tau/T_0=1/1500$, $W/T_0=1/1500$, and $d=2$,
we find $\Gamma=0.031,0.0011,0.00085,0.0011,0.031$
for $\alpha-\beta=0,\pi/8,\pi/4,3\pi/8,\pi/2$, respectively.

\begin{figure*}[t]
\begin{center}
\mbox{
\includegraphics[width=8cm]{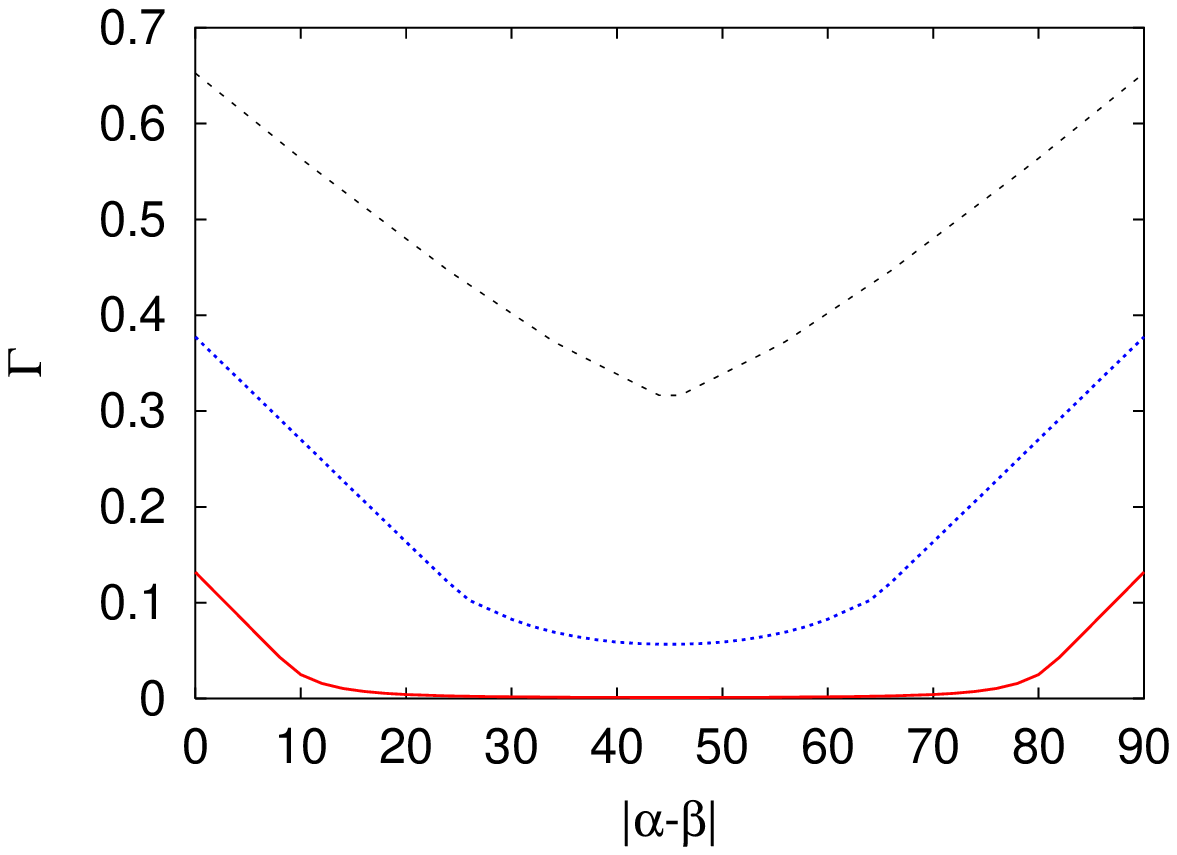}
\includegraphics[width=8cm]{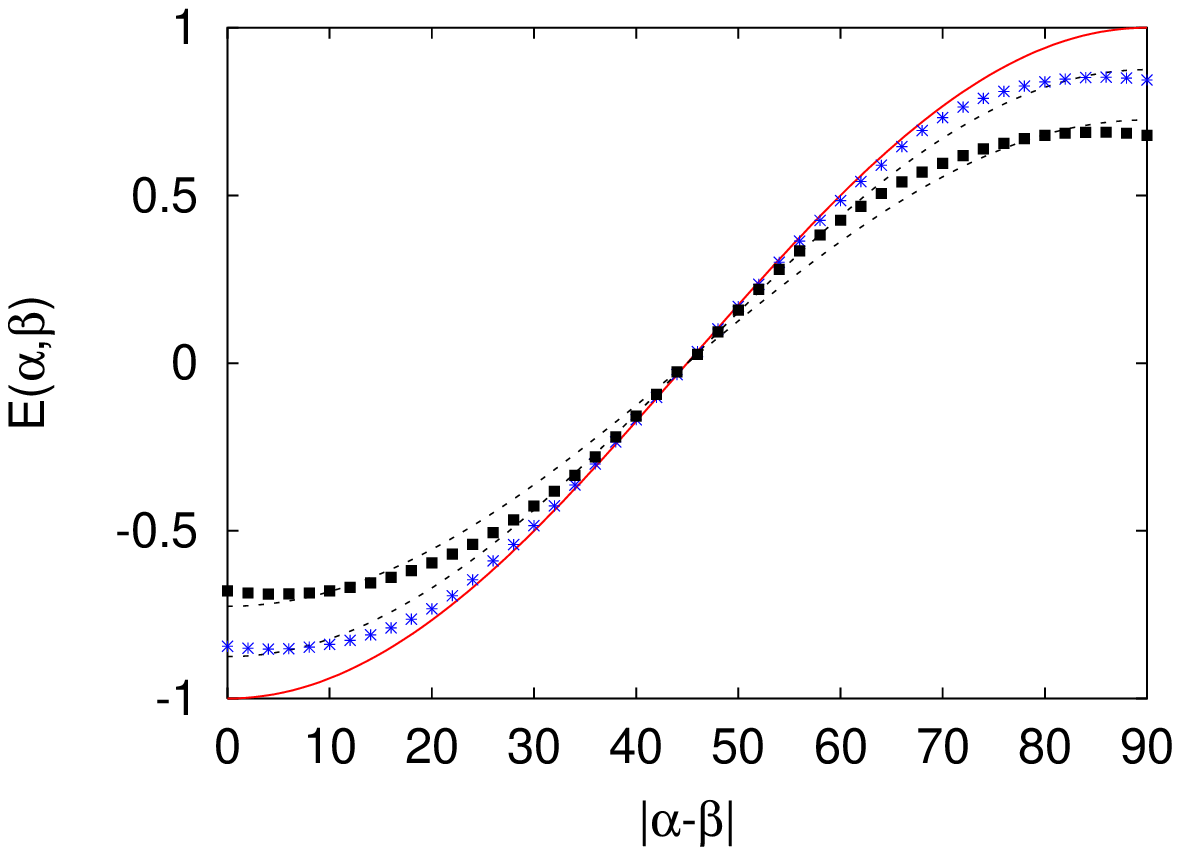}
}
\caption{(color online)
Left: The frequency of coincidences $\Gamma$ as a function of $|\alpha-\beta|$
for parameters $\tau/T_0$, $W/T_0$ and $d$ chosen such (see text)
that the simulation model reproduces quantum theory (solid line, red),
$S_{max}=2.83$, and the values of
$S_{max}=2.25$ (dashed line, black) and
$S_{max}=2.73$ (dotted line, blue),
as obtained from experiments
with ions~\cite{ROWE01}
and
with photons~\cite{WEIH98}, respectively.
Right:
Simulation results of the two-particle correlation $E(\alpha-\beta)$ as a function of $|\alpha-\beta|$
for the model parameters that yield $S_{max}=2.25$ (squares, black) and
$S_{max}=2.73$ (stars, blue), respectively.
The dashed lines are given by
$-0.875\cos2(\alpha-\beta)$
and
$-0.725\cos2(\alpha-\beta)$.
The solid line (red) is the result $-\cos2(\alpha-\beta)$ of quantum theory (see Fig.~\ref{fig10} (left)).
}
\label{gamma}
\label{gamma2}
\end{center}
\end{figure*}

In Fig.~\ref{gamma}(left), we plot $\Gamma$ as a function of $\alpha-\beta$, as obtained for the
simulation model introduced in the paper, for the three cases $S_{max}=2.25,2.73,2.83$ discussed earlier.
The general trend is clear: $\Gamma$ reaches its maximum at $\alpha-\beta=0,\pi/2,\ldots$
and its (nonzero) minimum at $\alpha-\beta=\pi/4,3\pi/2,\ldots$.

Finally, we study how $E(\alpha-\beta)$ deviates from the
result $E(\alpha-\beta)=-\cos2(\alpha-\beta)$ of a system in the singlet state as we fit
the values of $S_{max}$ to the experimental results.
In Fig.~\ref{gamma2}(right), we show the simulation results for the two cases $S_{max}=2.25,2.73$,
corresponding to the experiment
with ions of Ref.~\cite{ROWE01}
and the experiment
with photons of Ref.~\cite{WEIH98}, respectively.
From Fig.~\ref{gamma2}(right), we see that the main effect of reducing $S_{max}$ is
to reduce the amplitude (visibility) of the correlation.
Although it is clear that the simulation data
cannot be described by a single sinusoidal function, the deviations are small and it remains to be seen if experiments
can resolve such small differences.

As is evident from Fig.~\ref{maxStheta},
for $d>4$ our model yields the value for the singlet state $S_{max}=2\sqrt{2}$
without having to consider the regime of small $W$.
Thus, in order for an experiment and a model of the type considered in our paper
to reproduce the features of a quantum system
of two $S=1/2$ particles in the singlet state, it is not sufficient to
show that it can yield $S_{max}=2\sqrt{2}$ for some
choice of the parameters. As mentioned earlier, the singlet state is completely characterized by the
single- and two-particle expectation values. Hence, in order to make a comparison with the singlet state,
it is necessary to measure or compute these two quantities.

\section{Discussion}\label{sec6}
\label{sec:discussion}

We have presented a computer algorithm that simulates Aspect-type EPRB experiments.
In the simulation, the source produces particles with opposite but otherwise unpredictable polarization.
Each particle of a pair is analyzed in an observation station, consisting of a polarizer and two
detectors (Case I). Placing an additional polarizer in between the source and
each observation station changes the opposite, unpredictable polarization of the two
particles into a pair of fixed, but not necessarily opposite, polarizations (Case II).
The time-tag data of the detection events observed in both stations are used for pair
identification. Application of quantum theory to both types of experiments
yields the single-particle and two-particle expectation values that are characteristic for the singlet state (Case I)
and the product state (Case II).

The salient features of the simulation model are that:
\begin{itemize}
\item{Every essential component of the real laboratory experiment
(polarizers, detectors, time-tag logic, data analysis procedure)
has a counterpart in the algorithm.}
\item{Identical elements in the experimental setup are represented by identical algorithms. For instance,
to simulate Case I and II, we use the same algorithm to simulate the polarizers.
In particular, the algorithm that simulates the polarizer reproduces Malus law,
which is not essential to reproduce the quantum theoretical results of Case I, see the model
introduced in Ref.~\cite{RAED06c}.}
\item{It is event-based and strictly satisfies Einstein's criteria of local causality but it is not unique.}
\item{At any time, it allows free choice of the directions in which the polarization will be measured,
in contrast to laboratory experiments in which the polarizers in the observation stations
can take $2\times2$ directions only~\cite{WEIH98}.}
\item{It identifies pairs based on the time-tag of each detection event, using a time window $W$ and allowing for several
different procedures to define which two photons form a pair, just as in real laboratory experiments.}
\item{To first order in the time window $W$,
it reproduces exactly the single-particle averages and two-particle correlations of quantum theory for both Case I and II.}
\item{It provides information about the frequency of coincidences $\Gamma$.
In order to reproduce the results of the two-particle correlation as given by quantum theory for Case I, $\Gamma$ must be sufficiently small.
Values of $\Gamma$, corresponding to those found in EPRB laboratory experiments~\cite{WEIH98,ROWE01} can be reproduced also.
For these values of $\Gamma$, the two-particle correlation function deviates from the quantum theoretical
result but the deviations are small. In all cases, the simulation model reproduces the single-particle averages
as given by quantum theory.}
\end{itemize}

In our simulation model, the time-tag data are
a key element for producing the single-particle expectation values and two-particle correlations as given by
the quantum theory of a system of two $S=1/2$ spins.
In our model, in Case I, the two-particle correlation depends on the value of the time window $W$.
By reducing $W$ from infinity to zero, this correlation changes from typical Bell-like to singlet-like,
{\sl without making any change to the whole algorithm}.
Thus, the character of the correlation not only depends on the whole experimental setup
but also on the way the data analysis is carried out.
Hence, from the two-particle correlation itself, one cannot make any definite statement about the character of the source.
Thus, the correlation is a property of the whole system
(which is what quantum theory describes), not a property of the source itself.
It is of interest to note that if we perform a simulation of Case II
the single-particle and two-particle correlations do not depend on the value of the time window $W$.
In this case, the observation stations always receive particles with the same polarization and
although the number of coincidences decreases with $W$ (and the statistical fluctuations increase),
the functional form of the correlation does not depend on $W$.

Summarizing: We have demonstrated that a simulation model that strictly satisfies Einstein's criteria of locality
can reproduce, event-by-event, the quantum theoretical results for EPRB experiments with photons,
without using any concept from quantum theory.
We have given a rigorous proof that this model reproduces the single-particle expectations and the two-particle
correlation of two $S=1/2$ particles in the singlet state and product state.

\section*{Acknowledgement}
We are grateful to R.D. Gill for pertinent comments on earlier versions of the manuscript
and for fruitful discussions and suggestions.


%
\bibliography{../../epr} 

\end{document}